\documentclass[prb,superscriptaddress,twocolumn,preprintnumbers,a4,showpacs,floatfix]{revtex4}
\usepackage{amsmath, amssymb, psfrag, tabls, dcolumn, bm, color}
\usepackage[sort&compress]{natbib}
\usepackage{graphicx}
\usepackage{textcomp}
\usepackage{multirow}
\usepackage{color}
\newcommand{\be}{\begin{equation}}
\newcommand{\ee}{\end{equation}}
\newcommand{\bea}{\begin{equation*}}
\newcommand{\eea}{\end{equation*}}
\newcommand{\ba}{\begin{array}}
\newcommand{\ea}{\end{array}}
\newcommand{\beqa}{\begin{eqnarray}}
\newcommand{\eeqa}{\end{eqnarray}}
\newcommand{\beqaa}{\begin{eqnarray*}}
\newcommand{\eeqaa}{\end{eqnarray*}}

\newcommand{\matr}{\left( \begin{array}}
\newcommand{\ematr}{\end{array} \right)}

\newcommand{\kb}{\mbox{\boldmath $k$}}

\newcommand{\rb}{\mbox{\boldmath $r$}}
\newcommand{\Rb}{\mbox{\boldmath $R$}}

\newcommand{\Fb}{\mbox{\boldmath $F$}}
\newcommand{\Lb}{\mbox{\boldmath $L$}}
\newcommand{\Tb}{\mbox{\boldmath $T$}}

\newcommand{\Vb}{\mbox{\boldmath $V$}}

\newcommand{\der}{{\rm d}}

\def\half{$\scriptstyle{1\over 2}$}

\newcommand{\lsim}{{\;\raise0.3ex\hbox{$<$\kern-0.75em\raise-1.1ex\hbox{$\sim$}}
\;}}
\newcommand{\gsim}{{\;\raise0.3ex\hbox{$>$\kern-0.75em\raise-1.1ex\hbox{$\sim$}}
\;}}

\newcommand{\dS}{\mbox{\boldmath $dS$}}
\newcommand{\dH}{\mbox{\boldmath $dH$}}

\usepackage{amssymb}

\newcommand{\equ}[1]{Eq.~(\ref{#1})}

\def\slantfrac#1#2{\kern.1em^{#1}\kern-.3em/\kern-.1em_{#2}}
\def\xx#1{\textrm{#1}}

\begin{document} 
\title{Density-Functional Tight-Binding for Beginners}
%
%
%

\author{Pekka Koskinen\footnote{Author to whom correspondence should be addressed.}}
\affiliation{NanoScience Center, Department of Physics, 40014 University of Jyv\"askyl\"a, Finland}
\email{pekka.koskinen@iki.fi}
 
\author{Ville M\"akinen}
\affiliation{NanoScience Center, Department of Physics, 40014 University of Jyv\"askyl\"a, Finland}


 
\date{\today}

\begin{abstract}
This article is a pedagogical introduction to density-functional tight-binding (DFTB) method. We derive it from the density-functional theory, give the details behind the tight-binding formalism, and give practical recipes for parametrization: how to calculate pseudo-atomic orbitals and matrix elements, and especially how to systematically fit the short-range repulsions. Our scope is neither to provide a historical review nor to make performance comparisons, but to give beginner's guide for this approximate, but in many ways invaluable, electronic structure simulation method---now freely available as an open-source software package, \texttt{hotbit}.
\end{abstract}


\maketitle

\section{Introduction}
If you were given only one method for electronic structure calculations, which method would you choose? It certainly depends on your field of research, but, on average, the usual choice is probably density-functional theory (DFT). It is not the best method for everything, but its efficiency and accuracy are suitable for most purposes. You may disagree with this argument, but already DFT community's size is a convincing evidence of DFT's importan\-ce---it is even among the few quantum mechanical methods used in industry. 

Things were different before. When computational resources were modest and DFT functionals inaccurate, classical force fields, semiempirical tight-binding, and jellium DFT calculations were used. Today tight-binding is mostly familiar from solid-state textbooks as a method for modeling band-structures, with one to several fitted hopping parameters\cite{castro_neto_RMP_09}. However, tight-binding could be used better than this more often even today---especially as a met\-hod to calculate total energies. Particularly useful for total energy calculations is density-functional tight-binding, which is parametrized directly using DFT, and is hence rooted in first principles deeper than other tight-binding flavors. It just happened that density-functional tight-binding came into existence some ten years ago\cite{porezag_PRB_95,elstner_PRB_98} when atomistic DFT calculations for realistic system sizes were already possible. With DFT as a competitor, the DFTB community never grew large.

Despite being superseded by DFT, DFTB is still useful in many ways: i) In calculations of large systems\cite{elstner_PSSb_00,frauenheim_PSSb_00}. Computational scaling of DFT limits system sizes, while better scaling is easier to achieve with DFTB. ii) Accessing longer time scales. Systems that are limited for optimization in DFT can be used for extensive studies on dynamical properties in DFTB\cite{koskinen_PRL_07}. iii) Structure search and general trends\cite{jackson_PRL_04}. Where DFT is limited to only a few systems, DFTB can be used to gather statistics and trends from structural families. It can be used also for pre-screening of systems for subsequent DFT calculations\cite{koskinen_NJP_06,koskinen_PRL_08}. iv) Method development. The formalism is akin to that of DFT, so methodology improvements, quick to test in DFTB, can be easily exported and extended into DFT\cite{bitzek_PRL_06}. v) Testing, playing around, learning and teaching. DFTB can be used simply for playing around, getting the feeling of the motion of atoms at a given temperature, and looking at the chemical bonding or realistic molecular wavefunctions, even with real-time simulations in a classroom---DFTB runs easily on a laptop computer. 

DFTB is evidently not an \emph{ab initio} method since it contains parameters, even though most of them have a theoretically solid basis. With parameters in the \emph{right} place, however, computational effort can be reduced enormously while maintaining a reasonable accuracy. This is why DFTB compares well with full DFT with minimal basis, for instance. Semiempirical tight-binding can be accurately fitted for a given test set, but transferability is usually worse; for general purposes DFTB is a good choice among the different tight-binding flavors.

Despite having its origin in DFT, one has to bear in mind that DFTB is still a tight-binding method, and should not generally be considered to have the accuracy of full DFT. Absolute transferability can never be achieved, as the fundamental starting point is \emph{tightly bound} electrons, with interactions ultimately treated perturbatively. DFTB is hence ideally suited for covalent systems such as hydro\-car\-bons.\cite{porezag_PRB_95,malola_PRB_08} Nevertheless, it does perform surprisingly well describing also metallic bonding with delocalized valence electrons\cite{kohler_CP_05,koskinen_NJP_06}.

This article is neither a historical review of different flavors of tight-binding, nor a review of even DFTB and its successes or failures. For these purposes there are several good reviews, see, for example Refs.~\onlinecite{frauenheim_JPCM_02,seifert_JCPA_07,otte_JCPA_07,elstner_JPCA_07}. Some ideas were around before\cite{seifert_ZPC_86,foulkes_PRB_89,sankey_PRB_89}, but the DFTB method in its present formulation, presented also in this article, was developed in the mid-90's\cite{porezag_PRB_95,frauenheim_PRB_95,seifert_IJQC_96,widany_PRB_96,elstner_PRB_98,liu_PRB_95}. The main architects behind the machinery were Eschrig, Seifert, Frauenheim, and their co-workers. We apologize for omitting other contributors---consult Ref.~\onlinecite{frauenheim_JPCM_02} for a more organized literature review on DFTB. 

Instead of reviewing, we intend to present DFTB in a pedagogical fashion. By occasionally being more explicit than usual, we derive the approximate formalism from DFT in a systematic manner, with modest referencing to how the same approximations were done before. The approach is practical: we present how to turn DFT into a \emph{working} tight-binding scheme where parametrizations are obtained from well-defined procedures, to yield actual numbers---without omitting ugly details hiding behind the scenes. Only basic quantum mechanics and selected concepts from density-functional theory are required as pre-requisite.

The DFTB parametrization process is usually presented as superficially easy, while actually it is difficult, especially regarding the fitting of short-range repulsion (Sec.~\ref{sec:fitting}). In this article we want to present a systematic scheme for fitting the repulsion, in order to accurately document the way the parametrization is done.

We discuss the details behind tight-binding formalism, like Slater-Koster integrals and transformations, largely unfamiliar for density-functional community; some readers may prefer to skip these detailed appendices. Because one of DFTB's strengths is the transparent electronic structure, in the end we also present selected analysis tools.

We concentrate on ground-state DFTB, leaving time-dependent\cite{todorov_JPCM_01,torralva_PRB_01,niehaus_EPJD_05,reimers_JPCA_07} or linear response\cite{niehaus_PRB_01} formalisms outside the discussion. We do not include spin in the formalism. Our philosophy lies in the limited benefits of improving upon spin-paired self-consistent -charge DFTB. Admittedly, one can adjust parametrizations for certain systems, but the tight-binding formalism, especially the presence of the repulsive potential, contains so many approximations that the next level in accuracy, in our opinion, is full DFT.

This philosophy underlies \texttt{hotbit}\cite{hotbit_wiki} software. It is an open-source DFTB package, released under the terms of GNU general public license\cite{GPL}. It has an interface with the atomic simulation environment (ASE)\cite{ase_wiki}, a python module for multi-purpose atomistic simulations. The ASE interface enables simulations with different levels of theory, including many DFT codes or classical potentials, with DFTB being the lowest-level quantum-mechanical method. \texttt{hotbit} is built upon the theoretical basis described here---but we avoid technical issues related practical implementations having no scientific relevance.

\section{The Origins of DFTB}

\subsection{Warm-up}
We begin by commenting on practical matters. The equations use $\hbar^2/m_e=4\pi \varepsilon_0=e=1$. This gives Bohr radius as the unit of length ($a_B=0.5292$~\AA) and Hartree as the unit of energy (Ha$=27.2114$~eV). Selecting atomic mass unit ($u=1.6605\cdot 10^{-27}$~kg) the unit of mass, the unit of time becomes $1.0327$~fs, appropriate for molecular dynamics simulations. Some useful fundamental constants are then $\hbar=\sqrt{m_e/u}=0.0234$, $k_B=3.1668\cdot 10^{-6}$, and $\varepsilon_0=1/(4\pi)$, for instance.

Electronic eigenstates are denoted by $\psi_a$, and (pseudo-atomic) basis states $\varphi_\mu$, occasionally adopting Dirac's notation. Greek letters $\mu$, $\nu$ are indices for basis states, while capital Roman letters $I$, $J$ are indices for atoms; notation $\mu \in I$ stands for orbital $\mu$ that belongs to atom $I$. Capital $\Rb$ denotes nuclear positions, with the position of atom $I$ at $\Rb_I$, and displacements $R_{IJ}=|\Rb_{IJ}|=|\Rb_J-\Rb_I|$. Unit vectors are denoted by $\hat{\Rb}=\Rb/|\Rb|$.

In other parts our notation remains conventional; deviations are mentioned or made self-explanatory.

\subsection{Starting Point: Full DFT}
The derivation of DFTB from DFT has been presented several times; see, for example Refs.~\onlinecite{foulkes_PRB_89,frauenheim_JPCM_02} and \onlinecite{elstner_PRB_98}. We do not want to be redundant, but for completeness we derive the equations briefly; our emphasis is on the final expressions.

We start from the total energy expression of interacting electron system
\begin{equation}
 E=T+E_{\xx{ext}}+E_{\xx{ee}}+E_{II},
\end{equation}
where $T$ is the kinetic energy, $E_{\xx{ext}}$ the external interaction (including electron-ion interactions), $E_{ee}$ the electron-electron interaction, and $E_{II}$ ion-ion interaction energy. Here $E_{II}$ contains terms like $Z_I^v Z_J^v/|\Rb_I-\Rb_J|$, where $Z_I^v$ is the valence of the atom $I$, and other contributions from the core electrons. In density-functional theory the energy is a functional of the electron density $n(\rb)$, and for Kohn-Sham system of non-interacting electrons the energy can be written as
\begin{equation}
 E[n(\rb)]=T_s+E_{\xx{ext}}+E_{H}+E_{xc}+E_{II},
\end{equation}
where $T_s$ is the non-interacting kinetic energy, $E_H$ is the Hartree energy, and $E_{xc}=(T-T_s)+(E_{ee}-E_H)$ is the exchange-correlation ($xc$) energy, hiding all the difficult many-body effects. More explicitly,
\begin{align}
\begin{split}
E[n]=&\sum_a f_a \langle \psi_a|\left( -\frac{1}{2}\nabla^2+ V_{\xx{ext}} + \frac{1}{2}\int \frac{n(\rb')\der^3r'}{|\rb'-\rb|} \right)| \psi_a \rangle \\
&+ E_{xc}[n] + E_{II},
\end{split}
\end{align}
where $f_a \in [0,2]$ is the occupation of a single-particle state $\psi_a$ with energy $\varepsilon_a$, usually taken from the Fermi-function (with factor $2$ for spin)
\begin{equation}
f_a=f(\varepsilon_a)=2 \cdot [\exp (\varepsilon_a-\mu)/k_BT+1]^{-1}
\end{equation}
with chemical potential $\mu$ chosen such that $\sum_a f_a=$ number of electrons. The Hartree potential
\begin{equation}
V_H[n](\rb)=\int' \frac{n(\rb')}{|\rb'-\rb|}, 
\end{equation}
is a classical electrostatic potential from given $n(\rb)$; for brevity we will use the notation $\int \der^3r \rightarrow \int$, $\int \der^3r' \rightarrow \int'$, $n(\rb) \rightarrow n$, and $n(\rb') \rightarrow n'$. With this notation the Kohn-Sham DFT energy is, once more,
\begin{align}
\begin{split}
E[n]=&\sum_a f_a \langle \psi_a|\left( -\frac{1}{2} \nabla^2 + \int V_{\xx{ext}}(\rb) \right) |\psi_a \rangle\\
     +& \frac{1}{2} \int \int' \frac{n n'}{|\rb - \rb'|} + E_{xc}[n] + E_{II}.
\end{split}
\end{align}

So far everything is exact, but now we start approximating. Consider a system with density $n_0(\rb)$ that is composed of atomic densities, as if atoms in the system were free and neutral. Hence $n_0(\rb)$ contains (artificially) no charge transfer. The density $n_0(\rb)$ does not minimize the functional $E[n(\rb)]$, but neighbors the true minimizing density $n_{min}(\rb)=n_0(\rb)+\delta n_0(\rb)$, where $\delta n_0(\rb)$ is supposed to be small. Expanding $E[n]$ at $n_0(\rb)$ to second order in fluctuation $\delta n(\rb)$ the energy reads
\begin{align}
\begin{split}
\label{eq:dn_expansion}
E[\delta n]\approx&\sum_a f_a \langle \psi_a| -\frac{1}{2}\nabla^2 + V_\text{ext} + V_H[n_0] + V_{xc}[n_0]| \psi_a \rangle \\
&+\frac{1}{2} \int \int' \left ( \frac{\delta^2 E_{xc}[n_0]}{\delta n \delta n'} + \frac{1}{|\rb-\rb'|} \right ) \delta n \delta n' \\
&-\frac{1}{2} \int V_H[n_0](\rb)n_0(\rb) + E_{xc}[n_0] + E_{II} \\
&- \int V_{xc}[n_0](\rb) n_0(\rb),
\end{split}
\end{align}
while linear terms in $\delta n$ vanish. The first line in \equ{eq:dn_expansion} is the band-structure energy
\begin{equation}
E_{BS}[\delta n]= \sum_a f_a \langle \psi_a| H[n_0]| \psi_a \rangle,
\end{equation}
where the Hamiltonian $H^0=H[n_0]$ itself contains no charge transfer. The second line in \equ{eq:dn_expansion} is the energy from charge fluctuations, being mainly Coulomb interaction but containing also $xc$-contributions
\begin{equation} 
E_{\xx{coul}}[\delta n]=\frac{1}{2} \int \int' \left ( \frac{\delta^2 E_{xc}[n_0]}{\delta n \delta n'} + \frac{1}{|\rb-\rb'|} \right ) \delta n \delta n'.
\label{eq:ecoul}
\end{equation}
The third and fourth lines in \equ{eq:dn_expansion} are collectively called the repulsive energy
\begin{align}
\begin{split}
E_{\xx{rep}}=&-\frac{1}{2} \int V_H[n_0](\rb)n_0(\rb) + E_{xc}[n_0] + E_{II} \\
 &- \int V_{xc}[n_0](\rb) n_0(\rb),
\end{split}
\label{eq:repulsive_origin}
\end{align}
because of the ion-ion repulsion term. Using this terminology the energy is
\begin{equation}
E[\delta n]=E_{BS}[\delta n]+E_{\xx{coul}}[\delta n]+E_{\xx{rep}}.
\end{equation}
Before switching into tight-binding description, we discuss $E_\text{rep}$ and $E_{\xx{coul}}$ separately and introduce the main approximations.

\subsection{Repulsive Energy Term}
In \equ{eq:repulsive_origin} we lumped four terms together and referred them to as repulsive interaction. It contains the ion-ion interaction so it is repulsive (at least at small atomic distances), but it contains also $xc$-interactions, so it is a complicated object. At this point we adopt manners from DFT: we sweep the most difficult physics under the carpet. You may consider $E_{\xx{rep}}$ as practical equivalent to an $xc$-functional in DFT because it hides the cumbersome physics, while we approximate it with simple functions. 

For example, consider the total volumes in the first term, the Hartree term
\begin{equation}
-\frac{1}{2}\int \int \frac{n_0(\rb)n_0(\rb')}{|\rb-\rb'|}\der^3r \der^3r'
\end{equation}
divided into atomic volumes; the integral becomes a sum over atom pairs with terms depending on atomic numbers alone, since $n_0(\rb)$ depends on them. We can hence approximate it as a sum of terms over atom pairs, where each term depends only on elements and their distance, because $n_0(\rb)$ is spherically symmetric for free atoms. Similarly ion-ion repulsions 
\begin{equation}
\frac{Z_I^v Z_J^v}{|\Rb_J-\Rb_I|}=\frac{Z_I^v Z_J^v}{R_{IJ}}
\end{equation}
depend only on atomic numbers via their valence numbers $Z_I^v$. Using similar reasoning for the remaining terms the repulsive energy can be approximated as
\begin{equation}
E_{\xx{rep}}=\sum_{I<J} V_\textrm{rep}^{IJ}(R_{IJ}).
\end{equation}
For each pair of atoms $IJ$ we have a repulsive function $V_\textrm{rep}^{IJ}(R)$ depending only on atomic numbers. Note that $E_{\xx{rep}}$ contains also on-site contributions, not only the atoms' pair-wise interactions, but these depend only on $n_0(\rb)$ and shift the total energy by a constant. 

The pair-wise repulsive functions $V_\text{rep}^{IJ}(R)$ are obtained by fitting to high-level theoretical calculations; detailed description of the fitting process is discussed in \ref{sec:fitting}.

\subsection{Charge Fluctuation Term}
\label{subsec:fluctuation}
Let us first make a side road to recall some general concepts from atomic physics. Generally, the atom energy can be expressed as a function of $\Delta q$ extra electrons as\cite{parr_book_94} 
\begin{align}
\begin{split}
E(\Delta q) &\approx E_0 + \left (\frac{\partial E}{\partial \Delta q} \right )  \Delta q + \frac{1}{2} \left (\frac{\partial^2 E}{\partial \Delta q^2} \right )\Delta q^2 \\
&= E_0 - \chi \Delta q + \frac{1}{2} U \Delta q^2.
\end{split}
\label{eq:parr}
\end{align}
The (negative) slope of $E(\Delta q)$ at $\Delta q=0$ is given by the (positive) electronegativity, which is usually approximated as 
\begin{equation}
\chi \approx (IE+EA)/2,
\end{equation}
where $IE$ the ionization energy and $EA$ the electron affinity. The (upward) curvature of $E(\Delta q)$ is given by the Hubbard $U$, which is
\begin{equation}
U \approx IE-EA,
\label{eq:U-defined}
\end{equation}
and is twice the atom absolute hardness $\eta$ ($U=2\eta$)\cite{parr_book_94}. Electronegativity comes mainly from orbital energies relative to the vacuum level, while curvature effects come mainly from Coulomb interactions.

Let us now return from our side road. The energy in \equ{eq:ecoul} comes from Coulomb and $xc$-interactions due to fluctuations $\delta n(\rb)$, and involves a double integrals over all space. Consider the space $\mathcal{V}$ divided into volumes $\mathcal{V}_I$ related to atoms $I$, such that 
\begin{equation}
\sum_I \mathcal{V}_I = \mathcal{V} \quad \text{and} \quad \int_\mathcal{V} = \sum_I \int_{\mathcal{V}_I}.
\end{equation}
We never precisely define what these volumes $\mathcal{V}_I$ exactly are---they are always used qualitatively, and the usage is case-specific. For example, volumes can be used to calculate the extra electron population on atom $I$ as
\begin{equation}
\Delta q_I \approx \int_{\mathcal{V}_I} \delta n(\rb) \der^3r.
\label{eq:atomic_charge}
\end{equation}
By using these populations we can decompose $\delta n$ into atomic contributions
\begin{equation}
\delta n(\rb) = \sum_I \Delta q_I \delta n_I(\rb),
\label{eq:dn_division}
\end{equation}
such that each $\delta n_I(\rb)$ is normalized, $\int_{\mathcal{V}_I} \delta n_I(r)\der^3r=1$. Note that Eqs.~(\ref{eq:atomic_charge}) and (\ref{eq:dn_division}) are internally consistent. Ultimately, this division is used to convert the double integral in \equ{eq:ecoul} into sum over atoms pairs $IJ$, and integrations over volumes $\int_{\mathcal{V}_I} \int_{\mathcal{V}_J}$.

First, terms with $I=J$ are
\begin{equation}
\frac{1}{2} \Delta q_I^2 \int_{\mathcal{V}_I} \int'_{\mathcal{V}_I} \left ( \frac{\delta^2 E_{xc}[n_0]}{\delta n \delta n'} + \frac{1}{|\rb-\rb'|} \right ) \delta n_I \delta n_I'.
\label{eq:full_fluctuation}
\end{equation}
Term depends quadratically on $\Delta q_I$ and by comparing to \equ{eq:parr} we can see that integral can be approximated by $U$. Hence, terms with $I=J$ become $\frac{1}{2} U_I \Delta q_I^2$.

\begin{figure}[t!]
\includegraphics[width=7cm]{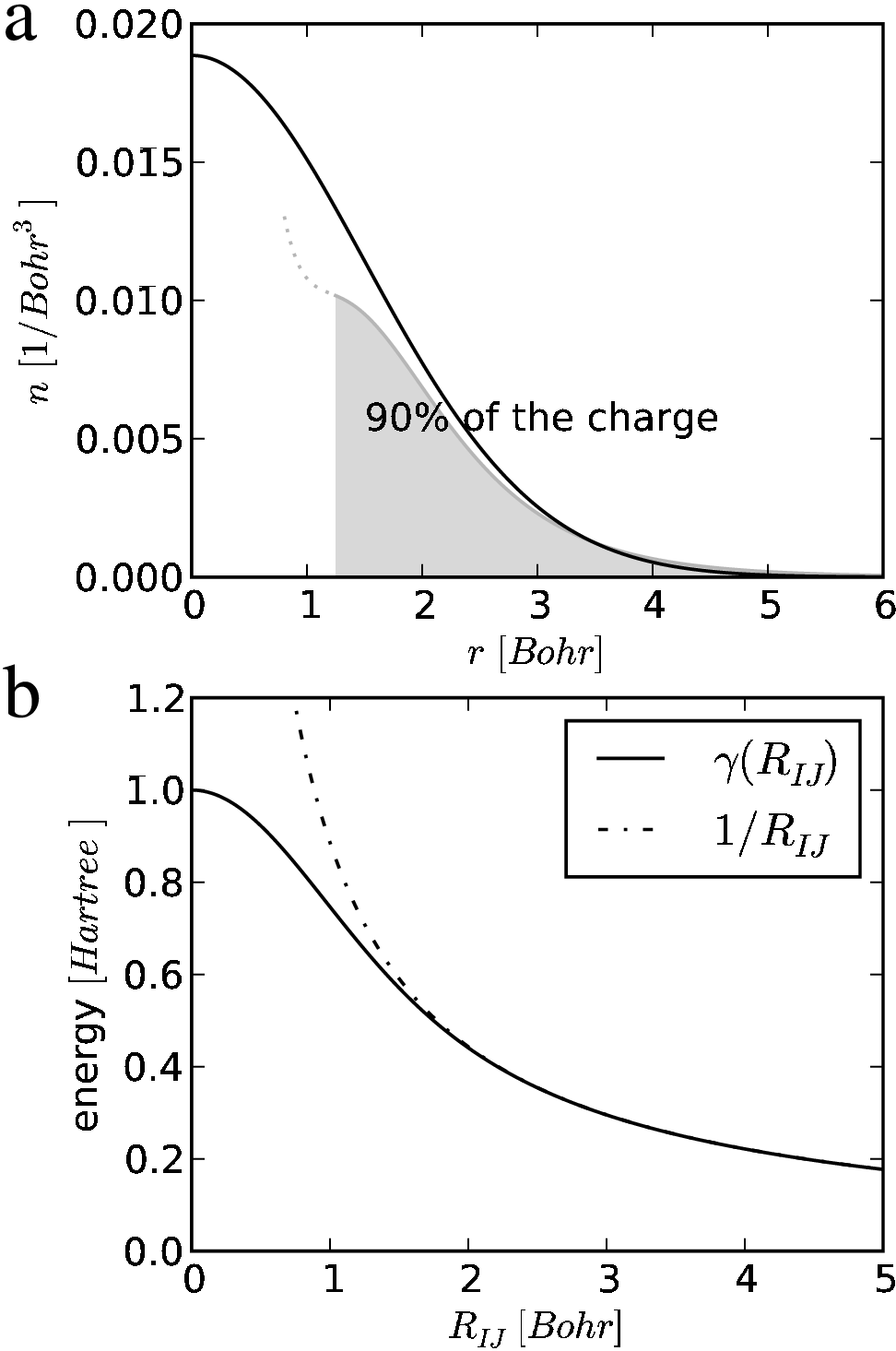}
\caption{(color online) (a) The change in density profile for C upon charging. Atom is slightly charged ($-2/3$~e and $+1$~e), and we plot the averaged $\delta n(r)=|n^\pm(r)-n_0(r)|$ (shadowed), where $n^\pm(r)$ is the radial electron density for charged atom, and $n_0(r)$ is the radial electron density for neutral atom. This is compared to the Gaussian profile of \equ{eq:gaussian-profile} with FWHM$=1.329/U$ where $U$ is given by \equ{eq:U-defined}. The change in density near the core is irregular, but the behavior is smooth for up to $\sim 90$~\% of the density change. (b) The interaction energy of two spherically symmetric Gaussian charge distributions with equal FWHM$_I=$FWHM$_J=1.329/U$ with $U=1$, as given by \equ{eq:gamma_function}. With $R_{IJ} \gg \textrm{FWHM}_I$ interaction is Coulomb-like, and approaches $U$ as $R_{IJ} \rightarrow 0$. }
\label{fig:delta_n-gamma-function}
\end{figure}

Second, when $I \neq J$ $xc$-contributions will vanish for local $xc$-functionals for which
\begin{equation}
\frac{\delta ^2 E_{xc}}{\delta n(\rb) \delta n(\rb')} \propto \delta(\rb-\rb'),
\end{equation}
and the interaction is only electrostatic,
\begin{equation}
\frac{1}{2} \Delta q_I \Delta q_J \int_{\mathcal{V}_I} \int'_{\mathcal{V}_J} \frac{\delta n_I \delta n_J'}{|\rb-\rb'|},
\label{eq:i_not_j}
\end{equation}
between extra atomic populations $\Delta q_I$ and $\Delta q_J$. Strictly speaking, we do not know what the functions $\delta n_I(\rb)$ are. However, assuming spherical symmetry, they tell how the density profile of a given atom changes upon charging. By assuming functional form for the profiles $\delta n_I(r)$, the integrals can be evaluated. We choose a Gaussian profile\cite{bernstein_PRB_02},
\begin{equation}
\delta n_I(r)=\frac{1}{(2\pi\sigma_I^2)^{3/2}} \exp \left (-\frac{r^2}{2\sigma_I^2} \right ),
\label{eq:gaussian-profile}
\end{equation}
where 
\begin{equation}
\sigma_I=\frac{\textrm{FWHM}_I}{\sqrt{8 \ln 2}}
\end{equation}
and FWHM$_I$ is the full width at half maximum for the profile. This choice of profile is justified for a carbon atom in Fig.~\ref{fig:delta_n-gamma-function}a. With these assumptions, the Coulomb energy of two spherically symmetric Gaussian charge distributions in \equ{eq:i_not_j} can be calculated analytically to yield
\begin{equation}
\int_{\mathcal{V}} \int'_{\mathcal{V}} \frac{\delta n_I \delta n_J'}{|\rb-\rb'|}=\frac{\textrm{erf}(C_{IJ}R_{IJ})} {R_{IJ}} \equiv \gamma_{IJ}(R_{IJ}),
\label{eq:gamma_function}
\end{equation}
where 
\begin{equation}
 C_{IJ}= \sqrt{\frac{4 \ln 2}{\textrm{FWHM}_I^2+\textrm{FWHM}_J^2}}.
\end{equation}
In definition (\ref{eq:gamma_function}) we integrate over whole spaces because $\delta n_I$'s are strongly localized. The function (\ref{eq:gamma_function}) is plotted in Fig.~\ref{fig:delta_n-gamma-function}b, where we see thata when $R \gg \xx{FWHM}$ we get point-like $1/R$-interaction. Furthermore, as $R \rightarrow 0$, $\gamma \rightarrow C\cdot 2/\sqrt{\pi}$, which gives us a connection to on-site interactions: if $I=J$
\begin{equation}
\gamma_{II}(R_{II}=0)=\sqrt{\frac{8 \ln 2}{\pi}}\frac{1}{\textrm{FWHM}_{I}}.
\label{eq:gamma_ii}
\end{equation}
This is the on-site Coulomb energy of extra population on atom $I$. Comparing to the $I=J$ case above, we can approximate
\begin{equation}
 \textrm{FWHM}_I=\sqrt{\frac{8 \ln 2}{\pi}} \frac{1}{U_I}=\frac{1.329}{U_I}.
\label{eq:FWHM_U}
\end{equation}
We interpret \equ{eq:gamma_ii} as: narrower atomic charge distributions causes larger costs to add or remove electrons; for a point charge charging energy diverges, as it should.

Hence, from absolute hardness, by assuming only Cou\-lom\-bic origin, we can estimate the sizes of the charge distributions, and these sizes can be used to estimate Coulomb interactions also between the atoms. $U$ and FWHM are coupled by \equ{eq:FWHM_U}, and hence for each element a single parameter $U_I$, which can be found from standard tables, determines all charge transfer energetics. 

To conclude this subsection, the charge fluctuation interactions can be written as
\begin{equation}
E_{\xx{coul}}=\frac{1}{2} \sum_{IJ} \gamma_{IJ}(R_{IJ})\Delta q_I \Delta q_J,
\label{eq:e_coul}
\end{equation}
where 
\begin{equation}
\gamma_{IJ}(R_{IJ})=
 \begin{cases}
 U_I, & I=J\\
 \frac{\textrm{erf}({C_{IJ}R_{IJ}})} {R_{IJ}}, & I \neq J.
\end{cases}
\end{equation}

\subsection{TB Formalism}
So far the discussion has been without references to tight-binding description. Things like eigenstates $|\psi_a \rangle$ or populations $\Delta q_I$ can be understood, but what are they exactly?

As mentioned above, we consider only valence electrons; the repulsive energy contains all the core electron effects. Since tight-binding assumes \emph{tightly bound} electrons, we use minimal local basis to expand
\begin{equation}
 \psi_a(\rb)=\sum_\mu c_\mu^a \varphi_\mu(\rb).
\label{eq:wfexpansion}
\end{equation}
Minimality means having only one radial function for each angular momentum state: one for $s$-states, three for $p$-states, five for $d$-states, and so on. We use real spherical harmonics familiar from chemistry---for completeness they are listed in Table~\ref{tab:spherical_functions} in Appendix~\ref{app:pseudo_atom}.

With this expansion the band-structure energy becomes
\begin{equation}
 E_{BS}=\sum_a f_a \sum_{\mu\nu} c_\mu^{a*} c_\nu^a H^0_{\mu\nu},
\end{equation}
where 
\begin{equation}
H^0_{\mu\nu}=\langle \varphi_\mu |H^0| \varphi_\nu \rangle. 
\label{eq:H0_mel}
\end{equation}
The tight-binding formalism is adopted by accepting the matrix elements $H^0_{\mu\nu}$ themselves as the principal parameters of the method. This means that in tight-binding spirit the matrix elements $H^0_{\mu\nu}$ are just \emph{numbers}. Calculation of these matrix elements is discussed in Section~\ref{sec:mels}, with details left in Appendix~\ref{app:mels}.

\begin{figure}
\begin{center}
\includegraphics[width=4cm]{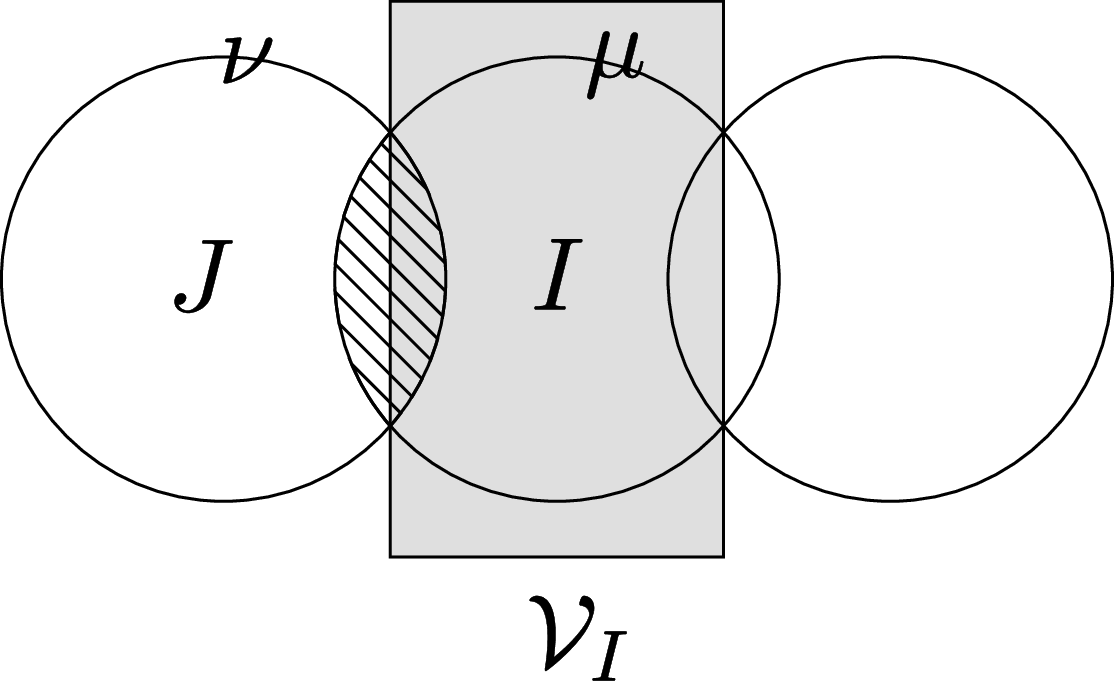}
\end{center}
\caption{Integrating the overlap of local orbitals. The large shaded area denotes the volume $\mathcal{V}_I$ of atom $I$, spheres represent schematically the spatial extent of orbitals, and the small hatched area denotes the overlap region. With $\nu \in J$ and $\mu \in I$, the integration of $\varphi_\mu(\rb)^*\varphi_\nu(\rb)$ over atom $I$'s volume $\mathcal{V}_I$ misses half of the overlap, since the other half is left approximately to $\mathcal{V}_J$.}
\label{fig:mulliken}
\end{figure}

How about the atomic populations $\Delta q_I$? Using the localized basis, the total number of electrons on atom $I$ is 
\begin{align}
\begin{split}
 q_I&=\sum_a f_a \int_{\mathcal{V}_I} |\psi_a(\rb)|^2 \der^3r\\
&=\sum_a f_a \sum_{\mu\nu} c_\mu^{a*} c_\nu^a \int_{\mathcal{V} _I} \varphi_\mu^*(\rb) \varphi_\nu(\rb)\der^3r. 
\end{split}
\end{align}
If neither $\mu$ nor $\nu$ belong to $I$, the integral is roughly zero, and if both $\mu$ and $\nu$ belong to $I$, the integral is approximately $\delta_{\mu\nu}$ since orbitals on the same atom are orthonormal. If $\mu$ belongs to $I$ and $\nu$ to some other atom $J$, the integral becomes  
\begin{equation}
\int_{\mathcal{V} _I} \varphi_\mu^*(\rb) \varphi_\nu(\rb) \approx \frac{1}{2}\int_{\mathcal{V}} \varphi_\mu^*(\rb) \varphi_\nu(\rb)=\frac{1}{2}S_{\mu\nu},
\end{equation}
as suggested by Fig.~\ref{fig:mulliken}, where $S_{\mu\nu}=\langle \varphi_\mu|\varphi_\nu \rangle$ is the overlap of orbitals $\mu$ and $\nu$. Charge on atom $I$ is
\begin{equation}
 q_I=\sum_a f_a \sum_{\mu \in I}\sum_\nu \frac{1}{2}(c_\mu^{a*}c_\nu^a + \textrm{c.c.})S_{\mu\nu},
\label{eq:mulliken_charges}
\end{equation}
where c.c. stands for complex conjugate. Hence $\Delta q_I=q_I-q_I^0$, where $q_I^0$ is the number of valence electrons for a neutral atom. This approach is called the Mulliken population analysis\cite{mulliken_JCP_55}.

Now were are ready for the final energy expression,
\begin{align}
\begin{split}
 E=&\sum_a f_a \sum_{\mu\nu} c_\mu^{a*}c_\nu^a H_{\mu\nu}^0 \\
&+ \frac{1}{2}\sum_{IJ} \gamma_{IJ}(R_{IJ})\Delta q_I \Delta q_J + \sum_{I<J} V_{\xx{rep}}^{IJ}(R_{IJ}),
\end{split}
\label{eq:energy_expression}
\end{align}
where everything is, in principle, defined. We find the minimum of this expression by variation of $\delta(E-\sum_a \varepsilon_a \langle \psi_a|\psi_a \rangle)$, where $\varepsilon_a$ are undetermined Lagrange multipliers, constraining the wave function norms, and obtain
\begin{equation}
\sum_\nu c_\nu^a (H_{\mu\nu}-\varepsilon_a S_{\mu\nu})=0,
\label{eq:kohn-sham}
\end{equation}
for all $a$ and $\mu$. This equation for the coefficients $c_\mu^a$ is the Kohn-Sham equation -equivalent in DFTB. Here
\begin{equation}
 H_{\mu\nu}=H^0_{\mu\nu} + \frac{1}{2}S_{\mu\nu}\sum_K (\gamma_{IK}+\gamma_{JK})\Delta q_K,\;\; \mu \in I\;\; \nu \in J.
\label{eq:Hmunu}
\end{equation}
By noting that the electrostatic potential on atom $I$ due to charge fluctuations is $\epsilon_I=\sum_K \gamma_{IK}\Delta q_K$, the equation above can be written as
\begin{equation}
 H_{\mu\nu}=H^0_{\mu\nu} + h^1_{\mu\nu}S_{\mu\nu},
\end{equation}
where
\begin{equation}
h^1_{\mu\nu}=\frac{1}{2} (\epsilon_I + \epsilon_J),\;\; \mu \in I\;\; \nu \in J.
\label{eq:h1_munu}
\end{equation}
This expression suggests a reasonable interpretation: charge fluctuations shift the matrix element $H_{\mu\nu}$ according to the averaged electrostatic potentials around orbitals $\mu$ and $\nu$. As in Kohn-Sham equations in DFT, also Eqs.~(\ref{eq:kohn-sham}) and (\ref{eq:Hmunu}) have to be solved self-consistently: from a given initial guess for $\{\Delta q_I\}$ one obtains $h_{\mu\nu}^1$ and $H_{\mu\nu}$, then by solving \equ{eq:kohn-sham} one obtains new $\{c_\mu^a\}$, and, finally, new $\{\Delta q_I\}$, iterating until self-consistency is achieved. The number of iterations required for convergence is usually markedly less than in DFT, albeit similar convergence problems are shared.

Atomic forces can be obtained directly by taking gradients of \equ{eq:energy_expression} with respect to coordinates (parameters) $\Rb_I$. We get (with $\nabla_J=\partial/\partial \Rb_J$)
\begin{align}
 \Fb_I=&-\sum_a f_a \sum_{\mu\nu} c_\mu^{a*} c_\nu^a \left [ \nabla_I H^0_{\mu\nu} - (\varepsilon_a-h^1_{\mu\nu})\nabla_I S_{\mu\nu} \right ] \nonumber \\
-& \Delta q_I \sum_J (\nabla_I \gamma_{IJ})\Delta q_J - \nabla_I E_{rep},
\label{eq:force}
\end{align}
where the gradients of $\gamma_{IJ}$ are obtained analytically from \equ{eq:gamma_function}, and the gradients of $H^0_{\mu\nu}$ and $S_{\mu\nu}$ are obtained numerically from an interpolation, as discussed in Appendix~\ref{app:mels}.

\section{Matrix Elements}
\label{sec:mels}
Now we discuss how to calculate the matrix elements $H^0_{\mu\nu}$ and $S_{\mu\nu}$. In the main text we describe only main ideas; more detailed issues are left to Appendices~\ref{app:pseudo_atom}, \ref{app:slako}, and \ref{app:mels}.

\subsection{The Pseudo-atom} 
The minimal basis functions $\varphi_\mu$ in the expansion (\ref{eq:wfexpansion}), 
\begin{equation}
\varphi_\mu(\rb')=R_\mu(r)\tilde{Y}_\mu(\theta,\varphi)\indent (\rb'=\Rb_I+\rb,\;\mu \in I),
\end{equation}
with real spherical functions $\tilde{Y}_\mu(\theta,\varphi)$ as defined in Appendix~\ref{app:pseudo_atom}, should robustly represent bound electrons in a solid or molecule, which is what we ultimately want to simulate. Therefore orbitals should not come from free atoms, as they would be too diffuse. To this end, we use the orbitals from a \emph{pseudo-atom}, where an additional confinement potential $V_\textrm{conf}(r)$ is added to the Hamiltonian
\begin{equation}
-\frac{1}{2}\nabla^2 - \frac{Z}{r} + V_H(r) + V_{xc}(r) + V_\textrm{conf}(r).
\end{equation}
This additional, spherically symmetric confinement cuts the orbitals' diffuse tails off and makes a compact basis---and ultimately better basis\cite{seifert_ZPC_86}---for the wave function expansion.

A general, spherically symmetric environment can be represented by a potential
\begin{equation}
V_\textrm{conf}(r)=\sum_{i=0}^\infty v_{2i} r^{2i},
\end{equation}
where the odd terms disappear because the potential has to be smooth at $r=0$. Since the first $v_0$ term is just a constant shift, the first non-trivial term is $v_2 r^2$. To first approximation we hence choose the confining potential to be of the form
\begin{equation}
V_\textrm{conf}(r)=\left ( \frac{r}{r_0} \right )^2,
\end{equation}
where $r_0$ is a parameter. The quadratic form for the confinement has appeared before\cite{porezag_PRB_95}, but also other forms have been analyzed\cite{junquera_PRB_01}. While different forms can be considered for practical reasons, they have only little effect on DFTB performance. The adjustment of the parameter $r_0$ is discussed in Section~\ref{sec:summary}.

The pseudo-atom is calculated with DFT only once for a given confining potential. This way we get $\varphi_\mu$'s (more precisely, $R_\mu$'s), the localized basis functions, for later use in matrix element calculations. 

One technical detail we want to point out here concerns orbital conventions. Namely, once the orbitals $\varphi_\mu$ are calculated, their sign and other conventions \emph{should never change}. Fixed convention should be used in all simultaneously used Slater-Koster tables; using different conventions for same elements gives inconsistent tables that are plain nonsense. The details of our conventions, along with other technical details of the pseudo-atom calculations, are discussed in Appendix~\ref{app:pseudo_atom}.

\subsection{Overlap Matrix Elements}
Using the orbitals from pseudo-atom calculations, we need to calculate the overlap matrix elements
\begin{equation}
 S_{\mu\nu}=\int \varphi_\mu(\rb)^*\varphi_\nu(\rb) \der^3r.
\label{eq:s-mel}
\end{equation}
Since orbitals are chosen real, the overlap matrix is real and symmetric. 

The integral with $\varphi_\mu$ at $\Rb_I$ and $\varphi_\nu$ at $\Rb_J$ can be calculated also with $\varphi_\mu$ at the origin and $\varphi_\nu$ at $\Rb_{IJ}$. Overlap will hence depend on $\Rb_{IJ}$, or equivalently, on $R_{IJ}$ and $\hat{\Rb}_{IJ}$ separately. Fortunately, the dependence on $\hat{\Rb}_{IJ}$ is fully governed by \emph{Slater-Koster transformation rules}\cite{slater_PR_54}. Only one to three \emph{Slater-Koster integrals}, depending on the angular momenta of $\varphi_\mu$ and $\varphi_\nu$, are needed to calculate the integral with any $\hat{\Rb}_{IJ}$ for fixed $R_{IJ}$. These rules originate from the properties of spherical harmonics. 

The procedure is hence the following: we integrate numerically the required Slater-Koster integrals for a set of $R_{IJ}$, and store them in a table. This is done once for all orbital pairs. Then, for a given orbital pair, we interpolate this table for $R_{IJ}$, and use the Slater-Koster rules to get the overlap with any geometry---fast and accurately.

Readers unfamiliar with the Slater-Koster transformations can read the detailed discussion in Appendix~\ref{app:slako}. The numerical integration of the integrals is discussed in Appendix~\ref{app:mels}.

Before concluding this subsection, we make few remarks about non-orthogonality. In DFTB it originates naturally and inevitably from \equ{eq:s-mel}, because non-over\-lap\-ping orbitals with diagonal overlap matrix would yield also diagonal Hamiltonian matrix, which would mean chemically non-interacting system. The transferability of a tight-binding model is often attributed to non-orthogonality, because it accounts for the spatial nature of the orbitals more realistically. 

Non-orthogonality requires solving a generalized eigenvalue problem, which is more demanding than normal eigenvalue problem. Non-orthogonality complicates, for instance, also gauge transformations, because the phase from the transformations is not well defined for the orbitals due to overlap. The Peierls substitution\cite{pople_JCP_62}, while gauge invariant in orthogonal tight-binding\cite{boykin_PRBb_01,graf_PRB_95}, is not gauge invariant in non-orthogonal tight-binding (but affects only time-dependent formulation).

\subsection{Hamiltonian Matrix Elements}
From \equ{eq:H0_mel} the Hamiltonian matrix elements are
\begin{equation}
 H^0_{\mu\nu}=\int \varphi_\mu(\rb)^*\left ( -\frac{1}{2}\nabla^2 + V_s[n_0](\rb) \right) \varphi_\nu(\rb),
\end{equation}
where
\begin{equation}
V_s[n_0](\rb)=V_\text{ext}(\rb) + V_H[n_0](\rb) + V_{xc}[n_0](\rb)
\end{equation}
is the effective potential evaluated at the (artificial) neutral density $n_0(\rb)$ of the system. The density $n_0$ is determined by the atoms in the system, and the above matrix element between basis states $\mu$ and $\nu$, in principle, depends on the positions of all atoms. However, since the integrand is a product of factors with three centers, two wave functions and one potential (and kinetic), all of which are non-zero in small spatial regions only, reasonable approximations can be made.

First, for diagonal elements $H_{\mu\mu}$ one can make a one-center approximation where the effective potential within volume $\mathcal{V}_I$ is
\begin{equation}
V_s[n_0](\rb) \approx V_{s,I}[n_{0,I}](\rb),
\end{equation}
where $\mu \in I$. This integral is approximately equal to the eigenenergies $\varepsilon_\mu$ of free atom orbitals. This is only approximately correct since the orbitals $\varphi_\mu$ are from the confined atom, but is a reasonable approximation that ensures the correct limit for free atoms.

Second, for off-diagonal elements we make the two-center approximation: if $\mu$ is localized around atom $I$ and $\nu$ is localized around atom $J$, the integrand is large when the potential is localized either around $I$ or $J$ as well; we assume that the crystal field contribution from other atoms, when the integrand has three different localized centers, is small. Using this approximation the effective potential within volume $\mathcal{V}_I+\mathcal{V}_J$ becomes
\begin{equation}
V_s[n_0](\rb) \approx V_{s,I}[n_{0,I}](\rb) + V_{s,J}[n_{0,J}](\rb),
\end{equation}
where $V_{s,I}[n_{0,I}](\rb)$ is the Kohn-Sham potential with the density of a neutral atom. The Hamiltonian matrix element is
\begin{equation}
 H^0_{\mu\nu}=\int \varphi_\mu(\rb)^*\left ( -\frac{1}{2}\nabla^2 + V_{s,I}(\rb) + V_{s,J}(\rb)\right) \varphi_\nu(\rb),
\label{eq:hamiltonian-mel}
\end{equation}
where $\mu \in I$ and $\nu \in J$. Prior to calculating the integral, we have to apply the Hamiltonian to $\varphi_\nu$. But in other respects the calculation is similar to overlap matrix elements: Slater-Koster transformations apply, and only a few integrals have to be calculated numerically for each pair of orbitals, and stored in tables for future reference. See Appendix~\ref{app:Hmels} for details of numerical integration of the Hamiltonian matrix elements.

\section{Fitting the repulsive potential}
\label{sec:fitting}

In this section we present a systematic approach to fit the repulsive functions $V_{\text{rep}}^{IJ}(R_{IJ})$ that appear in \equ{eq:energy_expression}---and a systematic way to describe the fitting. But first we discuss some difficulties related to the fitting process.

The first and straightforward way of fitting is simple: calculate dimer curve $E_{DFT}(R)$ for the element pair with DFT, require $E_{DFT}(R)=E_{DFTB}(R)$, and solve 
\begin{equation}
V_\text{rep}(R)=E_{DFT}(R)-[E_{BS}(R)+E_\text{coul}(R)].
\end{equation}
We could use also other symmetric structures with $N$ bonds having equal $R_{IJ}=R$, and require
\begin{equation}
N\cdot V_\textrm{rep}(R)=E_{DFT}(R)-[E_{BS}(R)+E_\text{coul}(R)].
\end{equation}
In practice, unfortunately, it does not work out. The approximations made in DFTB are too crude, and hence a single system is insufficient to provide a robust repulsion. As a result, fitting repulsive potentials is difficult task, and forms the most laborous part of parametrizing in DFTB. 

Let us compare things with DFT. As mentioned earlier, we tried to dump most of the difficult physics into the repulsive potential, and hence $V_\text{rep}$ in DFTB has practical similarity to $E_{xc}$ in DFT. In DFTB, however, we have to make a new repulsion for each pair of atoms, so the testing and fitting labor compared to DFT functionals is multifold. Because $xc$-functionals in DFT are well documented, DFT calculations of a reasonably documented article can be reproduced, whereas reproducing DFTB calculations is usually harder. Even if the repulsive functions are published, it would be a great advantage to be able to precisely describe the fitting process; repulsions could be more easily improved upon.

Our starting point is a set of DFT structures, with geometries $\Rb$, energies $E_\textrm{DFT}(\Rb)$, and forces $\Fb^\textrm{DFT}$ (zero for optimized structures). A natural approach would be to fit $V_\text{rep}$ so that energies $E_\textrm{DFTB}(\Rb)$ and forces $\Fb^\textrm{DFTB}$ are as close to DFT ones as possible. In other words, we want to minimize force differences $|\Fb^\textrm{DFT}-\Fb^\textrm{DFTB}|$ and energy differences $|E_\textrm{DFT}-E_\textrm{DFTB}|$ on average for the set of structures. There are also other properties such as basis set quality (large overlap with DFT and DFTB wave functions), energy spectrum (similarity of DFT and DFTB density of states), or charge transfer to be compared with DFT, but these originate already from the electronic part, and should be modified by adjusting $V_\text{conf}(r)$ and Hubbard $U$. Repulsion fitting is always the last step in the parametrizing, and affects only energies and forces.

In practice we shall minimize, however, only force diffe\-ren\-ces---we fit repulsion derivative, not repulsion directly. The fitting parameters, introduced shortly, can be adjusted to get energy differences qualitatively right, but only forces are used in the practical fitting algorithm. There are several reasons for this. First, forces are absolute, energies only relative. For instance, since we do not consider spin, it is ambiguous whether to fit to DFT dimer curve with spin-polarized or spin-paired free atom energies. We could think that lower-level spin-paired DFT is the best DFTB can do, so we compare to spin-paired dimer curve---but we should fit to energetics of nature, not energetics in some flavors of DFT. Second, for faithful dynamics it is necessary to have right forces and right geometries of local energy minima; it is more important first to get local properties right, and afterwards look how the global properties, such as energy ordering of different structural motifs, come out. Third, the energy in DFTB comes mostly from the band-structure part, not repulsion. This means that if already the band-structure part describes energy wrong, the short-ranged repulsions cannot make things right. For instance, if $E_\text{DFT}(R)$ and $E_\text{DFTB}(R)$ for dimer deviates already with large $R$, short-range repulsion cannot cure the energetics anymore. For transferability repulsion has to be monotonic and smooth, and if repulsion is adjusted too rapidly catch up with DFT energetics, the forces will go wrong.

For the set of DFT structures, we will hence minimize DFT and DFTB force differences, using the recipes below.

\subsection{Collecting Data}
To fit the derivative of the repulsion for element pair $AB$, we need a set of data points $\{R_i,V_\text{rep}'(R_i)\}$. As mentioned before, fitting to dimer curve alone does not give a robust repulsion, because the same curve is supposed to work in different chemical environments. Therefore it is necessary to collect the data points from several structures, to get a representative average over different types of chemical bonds. Here we present examples on how to acquire data points.

\subsubsection{Force Curves and Equilibrium Systems} 
This method can be applied to any system where all the bond lengths between the elements equal $R_{AB}$ or otherwise are beyond the selected cutoff radius $R_\text{cut}$. In other words, the only energy component missing from these systems is the repulsion from $N$ bonds between elements $A$ and $B$ with matching bond lengths. Hence, 
\begin{align}
\begin{split}
E_\text{DFTB}(R_{AB}) = \quad &E_{BS}(R_{AB}) + E_\text{coul}(R_{AB}) \\
+ &\tilde E_\text{rep} + N\cdot V_\text{rep}(R_{AB}),
\end{split}
\end{align}
where $\tilde E_\text{rep}$ is the repulsive energy independent of $R_{AB}$. This setup allows us to change $R_{AB}$, and we will require 
\be
V_\text{rep}'(R_{AB}) = \frac{E_\text{DFT}'(R_{AB})-[E_{BS}'(R_{AB})+E_\text{coul}'(R_{AB})]}{N},
\label{eq:curve_fitting}
\ee
where the prime stands for a derivative with respect to $R_{AB}$. The easiest way is first to calculate the energy curve and use finite differences for derivatives. In fact, systems treated this way can have even different $R_{AB}$'s if only the ones that are equal are chosen to vary (e.g. a complex system with one appropriate $AB$ bond on its surface). For each system, this gives a family of data points for the fitting; the number of points in the family does not affect fitting, as explained later. The dimer curve, with $N=1$, is clearly one system where this method can be applied. For any equilibrium DFT structure things simplify into
\begin{equation}
V_\text{rep}'(R_{AB}^0) = \frac{-E_{BS}'(R_{AB}^0)-E_\text{coul}'(R_{AB}^0)}{N},
\end{equation}
where $R_{AB}^0$ is the distance for which $E_\text{DFT}'(R_{AB}^0)=0$.

\subsubsection{Homonuclear Systems} 
If a cluster or a solid has different bond lengths, the energy curve method above cannot be applied (unless a subset of bonds are selected). But if the system is homonuclear, the data points can be obtained the following way. The force on atom $I$ is
\begin{align}
\Fb_I =& \Fb_I^0 + \sum_{J\neq I} V_\text{rep}'(R_{IJ}) \hat{R}_{IJ} \\
 =& \Fb_I^0 + \sum_{J\neq I} \epsilon_{IJ} \hat{R}_{IJ},
\end{align}
where $\Fb_I^0$ is the force without repulsions. Then we minimize the sum
\be
\sum_{I}\left|\Fb_{\text{DFT},I}-\Fb_I\right|^2
\ee
with respect to $\epsilon_{IJ}$, with $\epsilon_{IJ}=0$ for pair distances larger than the cutoff. The minimization gives optimum $\epsilon_{IJ}$, which can be used directly, together with their $R_{IJ}$'s, as another family of data points in the fitting.

\subsubsection{Other Algorithms}
Fitting algorithms like the ones above are easy to construct, but a few general guidelines are good to keep in mind.

While pseudo-atomic orbitals are calculated with LDA-DFT, the systems to fit the repulsive potential should be state-of-the-art calculations; all structural tendencies---whether right or wrong---are directly inherited by DFTB. Even reliable experimental structures can be used as fitting structures; there is no need to think DFTB should be parametrized only from theory. DFTB will not become any less \emph{density-functional} by doing so.

As data points are calculated by stretching selected bonds (or calculating static forces), also other bonds may stretch (dimer is one exception). These other bonds should be large enough to exclude repulsive interactions; otherwise fitting a repulsion between two elements may depend on repulsion between some other element pairs. While this is not illegal, the fitting process easily becomes complicated. Sometimes the stretching can affect chemical interactions between elements not involved in the fitting; this is worth avoiding, but sometimes it may be inevitable.

\subsection{Fitting the Repulsive Potential}
\label{subsec:fitting}
Transferability requires the repulsion to be short-ranged, and we choose a cutoff radius $R_\text{cut}$ for which $V_\text{rep}(R_\text{cut})=0$, and also $V_\text{rep}'(R_\text{cut})=0$ for continuous forces. $R_\text{cut}$ is one of the main parameters in the fitting process. Then, with given $R_\text{cut}$, after having collected enough data points $\{R_i,V_{\text{rep},i}'\}$, we can fit the function $V_\text{rep}'(R)$. The repulsion itself is 
\be
V_\text{rep}(R) = -\int^{R_\text{cut}}_R V_\text{rep}'(r)\der r.
\ee

\begin{figure}
\begin{center}
\includegraphics[width=7cm]{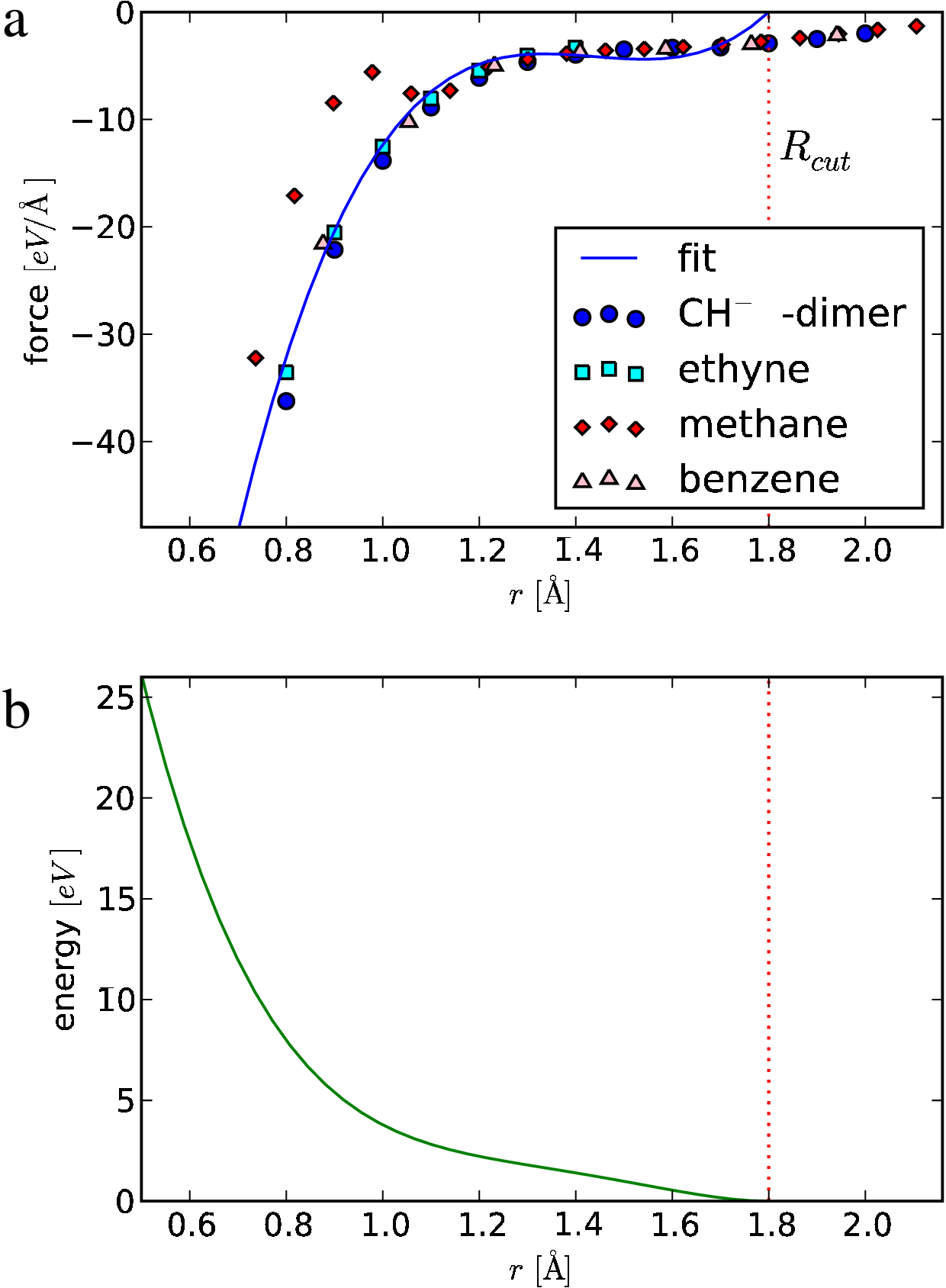}
\end{center}
\caption{(color online) a) Fitting the derivative of repulsive potential. Families of points from various structures, obtained by stretching C-H bonds and using \equ{eq:curve_fitting}. Here $R_\text{cut}=1.8$~\AA, for other details see Section~\ref{sec:summary}. b) The repulsive potential $V_\text{rep}(R)$, which is obtained from by integration of the curve in (a).}
\label{fig:fitting}
\end{figure}

Fitting of $V_\text{rep}'$ using the recipe below provides a robust and unbiased fit to the given set of points, and the process is easy to control. We choose a standard smoothing spline\cite{brohnstein_book_04} for $V_\text{rep}'(R)\equiv U(R)$, i.e. we minimize the functional
\be
S\left[U(R)\right] = \sum_{i=1}^{M} \left ( \frac{V_{\text{rep},i}' -U(R_i)}{\sigma_i} \right ) ^2 + \lambda\int^{R_\text{cut}} U''(R)^2 \der R
\ee
for total $M$ data points $\{R_i,V_{\text{rep},i}'\}$, where $U(R)$ is given by a cubic spline. Spline gives an unbiased representation for $U(R)$, and the smoothness can be directly controlled by the parameter $\lambda$. Large $\lambda$ means expensive curvature and results in linear $U(R)$ (quadratic $V_\text{rep}$) going through the data points only approximately, while small $\lambda$ considers curvature cheap and may result in a wiggled $U(R)$ passing through the data points exactly. The parameter $\lambda$ is the second parameter in the fitting process. Other choices for $U(R)$ can be used, such as low-order polynomials\cite{porezag_PRB_95}, but they sometimes behave surprisingly while continuously tuning $R_\text{cut}$. For transferability the behavior of the derivative should be as smooth as possible, preferably also monotonous (the example in Fig.~\ref{fig:fitting}a is slightly non-monotonous and should be improved upon).

The parameters $\sigma_i$ are the data point uncertainties, and can be used to weight systems differently. With the dimension of force, $\sigma_i$'s have also an intuitive meaning as force uncertainties, the lengths of force error bars. As described above, each system may produce a family of data points. We would like, however, the fitting to be independent of the number of points in each family; a fit with dimer force curve should yield the same result with $10$ or $100$ points in the curve. Hence for each system 
\begin{equation}
 \sigma_i=\sigma_s \sqrt{N_s},
\label{eq:sigma}
\end{equation}
where $\sigma_s$ is the uncertainty given for system $s$, with $N_s$ points in the family. This means that systems with the same $\sigma_s$'s have the same significance in the process, irregardless of the number of data points in each system. The effect of \equ{eq:sigma} is the same as putting a weight $1/N_s$ for each data point in the family. Note that $\lambda$ has nothing to do with the number of data points, and is more universal parameter. The cutoff is set by adding a data point at $U(R_\text{cut})=0$ with a tiny $\sigma$.

Fig.~\ref{fig:fitting} shows an example of fitting carbon-hydrogen repulsion. The parameters $R_\text{cut}$ and $\lambda$, as well as parameters $\sigma_i$, are in practice chosen to yield visually satisfying fit; the way of fitting should not affect the final result, and in this sense it is just a technical necessity---the simple objective is to get a smooth curve going nicely through the data points. Visualization of the data points can be also generally invaluable: deviation from a smooth behavior tells how well you should expect DFTB to perform. If data points lay nicely along one curve, DFTB performs probably well, but scattered data points suggest, for instance, the need for improvements in the electronic part. In the next section we discuss parameter adjustment further.

\section{Adjusting Parameters}
\label{sec:summary}
In this section we summarize the parameters, give practical instructions for their adjustment, and give a demonstration of their usage. The purpose is to give an overall picture of the selection of knobs to turn while adjusting parametrization.

Occasionally one finds published comments about the performance of DFTB. While DFTB shares flaws and failures characteristic to the method itself, it should be noted that DFTB parametrizations are even more diverse than DFT functionals. A website in Ref.~\onlinecite{dftb.org}, maintained by the original developers of the method, contains various sets of parametrizations. While these parametrizations are of good quality, they are not unique. Namely, there exists no automated way of parametrizing, so that a straightforward process would give all parameters definite values. This is not necessarily a bad thing, since some handwork in parametrizing also gives feeling what is to be expected in the future, how well parametrizations are expected to perform.

\subsection{Pseudo-Atoms}
The basis functions, and, consequently, the matrix elements are determined by the confinement potentials $V_\text{conf}(r)$ containing the parameters $r_0$ for each element. The value $r_0=2\cdot r_\textrm{cov}$, where $r_\text{cov}$ is the covalent radius, can be used as a rule of thumb\cite{porezag_PRB_95}. Since the covalent radius is a measure for binding range, it is plausible that the range for environmental confining potential should depend on this scale---the number $2$ in this rule is empirical. 

With this rule of thumb as a starting point, the quality of the basis functions can be inspected and $r_0$ adjusted by looking at (i) band-structure (for solids), (ii) densities of states (dimer or other simple molecules), or (iii) amount of data point scatter in repulsion fit (see Subsection~\ref{subsec:fitting}, especially Fig.~\ref{fig:fitting}). Systems with charge transfer should be avoided herein, since the properties listed above would depend on electrostatics as well, which complicates the process; adjustment of $r_0$ should be independent of electrostatics.

The inspections above are easiest to make with homonuclear interactions, even though heteronuclear interactions are more important for some elements, such as for hydrogen. Different chemical environments can affect the optimum value of $r_0$, but usually it is fixed for all interactions of a given element.

\subsection{Electrostatics}
Electrostatic energetics, as described in Subsection~\ref{subsec:fluctuation}, are determined by the Hubbard $U$ parameter, having the default value $U=IE-EA$. Since $U$ is a value for a free atom, while atoms in molecules are not free, it is permissible to adjust $U$ in order to improve (i) charge transfer, (ii) density of states, (iii) molecular ionization energies and electron affinities, or (iv) excitation spectra for selected systems. Since $U$ is an atomic property, one should beware when using several different elements in fitting---charge transfer depends on $U$'s of other elements as well.

\equ{eq:FWHM_U} relates $U$ and FWHM of given atom together. But since \equ{eq:full_fluctuation} contains, in principle, also $xc$-contribu\-ti\-ons, the relation can be relaxed, if necessary. Since FWHM affects only pair-interactions, it is better to adjust the interactions directly like
\begin{equation}
C_{IJ} \rightarrow C_{IJ}/x_{IJ},
\label{eq:xij}
\end{equation}
where $x_{IJ}$ (being close to one) effectively scales both FWHM$_I$ and FWHM$_J$. If atomic FWHMs would be changed directly, it would affect all interactions and complicate the adjusting process. Note that FWHM affects only nearest neighbor interactions (see Fig.~\ref{fig:delta_n-gamma-function}) and already next-nearest neighbors have (very closely) the pure $1/R$ interaction. 

An important principle, general for all parameters but particularly for electrostatics, is this: all parameters should be adjusted within reasonable limits. This means that, since all parameters have a physical meaning, if a parameter is adjusted beyond a reasonable and physically motivated limit, the parametrization will in general \emph{not be transferable}. If a good fit should require overly large parameter adjustments (precise ranges are hard to give), the original problem probably lies in the foundations of tight-binding.

\subsection{Repulsive Potentials}
The last step in the parametrization is to fit the repulsive potential. Any post-adjustment of other parameters calls for re-fitting of the repulsive potential. 

The most decisive part in the fitting is choosing the set of structures and bonds to fit. Parameters $R_\text{cut}$, $\sigma$ and $\lambda$ are necessary, but they play only a limited part in the quality of the fit---the quality and transferability is determined by band structure energy, electrostatic energy, and the chosen structures. In fact, the repulsion fitting was designed such that the user has as \emph{little} space for adjustment as possible.

The set of structures should contain the fitted interaction in different circumstances, with (i) different bond lengths, (ii) varying coordination, and (iii) varying charge transfer. In particular, if charge transfer is important for the systems of interest, calculation of charged molecules is recommended.

A reasonable initial guess for the cutoff radius is $R_\text{cut}=1.5\times R_\text{dimer}$, being half-way between nearest and next-nearest neighbors for homonuclear systems. It is then adjusted to yield a satisfying fit for the derivative of the repulsion, as in Fig.~\ref{fig:fitting}, while remembering that it has to be short ranged ($R_\text{cut}=2\times R_\text{dimer}$, for instance, is too large, lacks physical motivation, and makes fitting hard). Increasing $R_\text{cut}$ will increase $V_\text{rep}(R)$ at given $R<R_\text{cut}$, which is an aspect that can be used to adjust energies (but not much forces). Usually the parameters $\sigma$ are used in the sense of relative weights between systems, as often they cannot be determined in the sense of absolute force uncertainties. The absolute values do not even matter, since the scale of $\sigma$'s merely sets the scale for the smoothness parameter $\lambda$ (you can start with $\sigma=1$ for the first system; if you multiply $\sigma$'s by $x$, the same fit is obtained with $\lambda$ multiplied by $x^2$)---this is why the parameter $\lambda$ is not given any guidelines here. Large $\sigma$'s can be used to give less weight for systems with (i) marginal importance for systems of interest, (ii) inter-dependence on other parameters (dependence on other repulsions, on electrostatics, or on other chemical interactions), (iii) statistically peculiar sticking out from the other systems (reflecting situation that cannot be described by tight-binding or the urge to improve electronic part).

\subsection{Hydrocarbon Parametrization}
To demonstrate the usage of the parameters, we present the hydrocarbon parametrization used in this article (this was first shot fitting without extensive adjustment, but works reasonably well). The Hubbard $U$ is given by \equ{eq:U-defined} and FWHM by \equ{eq:FWHM_U}, both for hydrogen and carbon. The force curves have been calculated with GPAW\cite{mortensen_PRB_05,gpaw_wiki} using the PBE $xc$-functional\cite{perdew_PRL_96}.

Hydrogen: $r_0=1.08$, $U=0.395$. Carbon: $r_0=2.67$, $U=0.376$. Hydrogen-carbon repulsion: $R_\text{cut}=3.40$, $\lambda=35$, and systems with force curve: CH$^-$, ethyne, methane, benzene; all with $\sigma_i=1$. Carbon-carbon repulsion: $R_\text{cut}=3.80$, $\lambda=200$, and systems with force curve: C$_2$, CC$^{2-}$, C$_3$, C$_4^{2-}$ with $\sigma_i=1$, and C$_3^{2-}$ with $\sigma_i=0.3$.

\section{External fields}
Including external potentials to the formalism is straightforward: just add one more term in the Hamiltonian of \equ{eq:hamiltonian-mel}. Matrix element with external (scalar) potential becomes, with plausible approximations,
\begin{align}
H_{\mu\nu}&\rightarrow H_{\mu\nu} + \int \varphi_\mu^*(\rb)V_\text{ext}(\rb)\varphi_\nu(\rb) \der ^3 r \nonumber\\
&\approx H_{\mu\nu} + V_\text{ext}(\Rb_I)\int_{\mathcal{V}_I} \varphi_\mu^*\varphi_\nu + V_\text{ext}(\Rb_J)\int_{\mathcal{V}_J} \varphi_\mu^*\varphi_\nu \nonumber\\
&\approx H_{\mu\nu} + \frac{1}{2}\left ( V_\text{ext}^I+V_\text{ext}^J \right )S_{\mu\nu}
\end{align}
for a smoothly varying $V_\text{ext}(\rb)$. The electrostatic part in the Hamiltonian is
\begin{equation}
h^1_{\mu\nu}=\frac{1}{2}\left (\epsilon_I + \epsilon_J +  V_\text{ext}^I+V_\text{ext}^J \right),
\label{eq:external-potential}
\end{equation}
and naturally extends \equ{eq:h1_munu}.

\section{van der Waals Forces}
Accurate DFT xc-functionals, which automatically yield the $R^{-6}$ long range attractive van der Waals interactions, are notoriously hard to make\cite{rydberg_PRL_03}. Since DFT in other respects is accurate with short-range interactions, it would be wrong to add van der Waals interactions by hand---addition inevitably modifies short-range parts as well.

DFTB, on the other hand, is more approximate, and adding physically motivated terms by hand is easier. In fact, van der Waals forces in DFTB can conceptually be thought of as modifications of the repulsive potential. Since dispersion forces are due to $xc$-contributions, one can see that for neutral systems, where $\delta n(\rb)\equiv 0$, dispersion has to come from \equ{eq:repulsive_origin}. However, in practice it is better to leave $V_\text{rep}^{IJ}$'s short-ranged and add the dispersive forces as additional terms
\begin{equation}
E_\text{vdW}=-\sum_{I<J} f_{IJ}(R_{IJ}) \frac{C_6^{IJ}}{R_{IJ}^6}
\label{eq:vdW}
\end{equation}
in the total energy expression. Here $f(R)$ is a damping function with the properties
\begin{align}
f(R)=
\begin{cases}
 \approx 1, \indent &R \gtrsim R_0\\
 \approx 0, \indent &R \lesssim R_0,
\end{cases}
\end{align}
because the idea is to switch off van der Waals interactions for distances smaller than $R_0$, a characteristic distance where chemical interactions begin to emerge.

The $C_6$-parameters depend mainly on atomic polarizabilities and have nothing to do with DFTB formalism. Care is required to avoid large repulsive forces, coming from abrupt behavior in $f(R)$ near $R\approx R_0$, which could result in local energy minima. For a detailed descriptions about the $C_6$-parameters and the form of $f(R)$ we refer to original Refs.~\onlinecite{halgren_JACS_92} and \onlinecite{elstner_JCP_01}; in this section we merely demonstrate how straightforward it is, in principle, to include van der Waals forces in DFTB.

\section{Periodic Boundary Conditions}

\subsection{Bravais Lattices}
Calculation of isolated molecules with DFTB is straightforward, but implementation of periodic boundary conditions and calculation of electronic band-structures is also easy\cite{koskinen_physscr_03}. As mentioned in the introduction, this is usually the first encounter with tight-binding models for most physicists; our choice was to discuss periodic systems at later stage.

In a crystal periodic in translations $\Tb$, the wave functions have the Bloch form 
\begin{equation}
\psi_a(\kb,\rb)=e^{i\kb \cdot \rb} u_a(\kb,\rb),
\end{equation}
where $u_a(\kb,\rb)$ is function with crystal periodicity\cite{ashcroft_book_76}. This means that a wave function $\psi_a(\kb,\rb)$ changes by a phase $e^{i\kb \cdot \Tb}$ in translation $\Tb$. We define \emph{new} basis functions, not as localized orbitals anymore, but as Bloch waves extended throughout the whole crystal
\begin{equation} 
\varphi_{\mu}(\kb,\rb)=\frac{1}{\sqrt{N}} \sum_{\Tb} e^{i \kb \cdot \Tb} \varphi_\mu(\rb-\Tb),
\end{equation}
where $N$ is the (infinite) number of unit cells in the crystal. The eigenfunction \emph{ansatz}
\begin{equation}
\psi_{a}(\kb,\rb) = \sum_\mu c_{\mu}^a(\kb) \varphi_{\mu}(\kb,\rb)
\label{eq:wf_ansatz}
\end{equation}
is then also an extended Bloch wave, as required by Bloch theorem, because $\kb$ is the same for all basis states. Matrix elements in this new basis are
\begin{equation}
S_{\mu\nu}(\kb,\kb')=\delta(\kb-\kb') S_{\mu \nu}(\kb)
\end{equation}
and
\begin{equation}
H_{\mu\nu}(\kb,\kb')=\delta(\kb-\kb') H_{\mu \nu}(\kb),
\end{equation}
where
\begin{align}
\begin{split}
S_{\mu \nu}(\kb)&= \sum_{\Tb} e^{i \kb \cdot \Tb} \left ( \int \varphi_\mu^*(\rb) \varphi_\nu(\rb-\Tb) \right )\\
&\equiv\sum_{\Tb} e^{i \kb \cdot \Tb} S_{\mu \nu}(\Tb),
\end{split}
\end{align}
and similarly for $H$. Obviously the Hamiltonian conserves $\kb$---that is why $\kb$ labels the eigenstates in the first place. Note that the new basis functions are usually not normalized.

Inserting the trial wave function (\ref{eq:wf_ansatz}) into \equ{eq:energy_expression} and by using the variational principle we obtain the secular equation
\begin{equation}
\sum_\nu c^a_\nu(\kb) \left [ H_{\mu\nu}(\kb)-\varepsilon_a(\kb) S_{\mu\nu}(\kb) \right ] = 0,
\end{equation}
where
\begin{equation}
H_{\mu\nu}(\kb)=H_{\mu\nu}^0(\kb) + h_{\mu\nu}^1 S_{\mu\nu}(\kb)
\end{equation}
for each $\kb$-point from a chosen set, such as Monkhorst-Pack sampled\cite{monkhorst_PRB_76}. Above we have 
\begin{equation}
h_{\mu\nu}^1=\frac{1}{2}( \epsilon_I + \epsilon_J ) \quad \mu \in I,\; \nu \in J
\end{equation}
as in \equ{eq:h1_munu}, and Mulliken charges are extensions of \equ{eq:mulliken_charges},
\begin{equation}
 q_I=\sum_a \sum_{\kb} f_a(\kb) \sum_{\mu \in I,\nu} \frac{1}{2} \left[ c_\mu^{a *}(\kb) c_\nu^a(\kb) S_{\mu\nu}(\kb) + \text{c.c.} \right].
\end{equation}

The sum for the electrostatic energy per unit cell,
\begin{equation}
E_\textrm{coul}=\frac{1}{2} \sum_{IJ}^\textrm{unit cell} \sum_{\Tb} \gamma_{IJ}(\Rb_{IJ}-\Tb)\Delta q_I \Delta q_J,
\end{equation}
can be calculated with standard methods, such as Ewald summation\cite{frenkel_book_02}, and the repulsive part,
\begin{equation}
\sum_{I<J} V_\text{rep}^{IJ} (R_{IJ}) = \frac{1}{2}\sum_{IJ}^\text{unit cell} \sum_{\Tb} V_{\text{rep}}^{IJ}(\Rb_{IJ}-\Tb)
\end{equation}
is easy because repulsions are short-ranged (and $V_\text{rep}(0)=0$ is understood).

\subsection{General Symmetries}
Thanks to the transparent formalism of DFTB, it is easy to construct more flexible boundary conditions, such as the ``wedge boundary condition'' introduced in Ref.~\onlinecite{malola_PRB_08b}. This is one example of DFTB in method development. 

General triclinic unit cells are copied by translations, and DFT implementation is easy with plane waves, real-space grids or localized orbitals with fixed quantization axis. But if we allow the quantization axis of localized orbitals to be position-dependent, we can treat more general symmetries which have rotational symmetry\cite{liu_JPCM_06} or even combined rotational and translational (chiral) symmet\-ri\-es\cite{popov_NJP_04}.

The basic idea is to enforce the orbitals to have the the same symmetry as the system. This requires that basis functions not only depend on atom positions like 
\begin{equation}
\varphi_\mu(\rb-\Rb_\mu),
\end{equation}
as usual, but more generally like
\begin{equation}
D(\Rb_\mu)\varphi_\mu(\rb),
\end{equation}
where $D(\Rb_\mu)$ is an operator transforming the orbitals in any position-dependent manner, including both translations and rotations. The only requirement is that the orbitals are complete and orthonormal for a given angular momentum. If the quantization axes change, things become unfortunately messy. However, suitably defined basis orbitals yield well-defined Hamiltonian and overlap matrices, and enable simulations of systems like bent tubes or slabs, helical structures such as DNA, or a piece of spherical surface---with a greatly reduced number of atoms. Similar concepts are familiar from chemistry, where symmetry-adapted molecular orbitals are constructed from the atomic orbitals, and computational effort is hereby reduced\cite{atkins_book_00}. Detailed treatment of these general symmetries is a work in progress\cite{kit}.

\section{Density-matrix Formulation}
In this section we introduce DFTB using density-matrix formulation. We do this because not only does the formulation simplify expressions, but it also makes calculations faster in practice. This practical advantage comes because the density-matrix,
\begin{equation}
 \rho_{\mu\nu}=\sum_a f_a (c_{\mu}^ac_{\nu}^{a*})=\sum_a f_a (\rho_{\mu\nu}^{a})
\label{eq:density-matrix},
\end{equation}
contains a loop over eigenstates; quantities calculated with $\rho_{\mu\nu}$ simply avoid this extra loop. It has the properties
\begin{align}
\rho S \rho &= \rho \quad (\sim\text{idempotency)}\\
\rho_{\mu\nu}&=\rho_{\nu\mu}^* \quad (\rho=\rho^\dagger)\\
\mathrm{Tr}(\rho^a S)&=1 \quad \text{(eigenfunction normalization)}.
\end{align}
We define also the energy-weighted density-matrix
\begin{equation}
 \rho^e_{\mu\nu}=\sum_a \varepsilon_a f_a c_{\mu}^ac_{\nu}^{a*},
\label{eq:energy-density-matrix}
\end{equation}
and symmetrized density matrix
\begin{equation}
\tilde{\rho}_{\mu\nu}=\frac{1}{2}(\rho_{\mu\nu}+\rho_{\mu\nu}^*),
\end{equation}
which is symmetric and real. Using $\rho_{\mu\nu}$ we obtain simple expressions, for example, for 
\begin{align}
E_{BS}&=\text{Tr}(\rho H^0)\\
N_{el}&=\text{Tr}(\tilde{\rho}S)=\sum_\mu \sum_\nu \tilde{\rho}_{\mu\nu} S_{\nu\mu}, \\
q_I&=\text{Tr}_I (\tilde{\rho} S)=\sum_{\mu \in I} \sum_\nu \tilde{\rho}_{\mu\nu} S_{\nu\mu},
\end{align}
where $E_{BS}$ is the band-structure energy, $N_{el}$ is the total number of electrons, and $q_I$ is the Mulliken population on atom $I$. $\text{Tr}_I$ is partial trace over orbitals of atom $I$ alone.

It is practical to define also matrices' gradients. They do not directly relate to density matrix formulation, but equally simplify notation, and are useful in practical implementations. We define (with $\nabla_J=\partial/\partial \Rb_J$)
\begin{align}
\begin{split}
\dS_{\mu\nu}&=\nabla_J S_{\mu\nu} \indent  \quad \mu \in I,\; \nu \in J\\ 
&\equiv \int \varphi_\mu^*(\rb-\Rb_I)\nabla_J\varphi_\nu(\rb-\Rb_J) \\
&= \langle \varphi_\mu| \nabla_J \varphi_\nu \rangle,
\end{split}
\end{align}
and
\begin{align}
\begin{split}
\dH_{\mu\nu} &=\nabla_J H_{\mu\nu} \indent  \quad \mu \in I,\; \nu \in J\\ 
&=\langle \varphi_\mu| H | \nabla_J \varphi_\nu \rangle
\end{split}
\end{align}
with the properties $\dS_{\mu\nu}=-\dS_{\nu\mu}$ and $\dH_{\mu\nu}=-\dH_{\nu\mu}^*$. From these definitions we can calculate analytically, for instance, the time derivative of the overlap matrix for a system in motion
\begin{equation}
\dot{S}=[\dS,\Vb],
\end{equation}
with commutator $[A,B]$ and matrix $\Vb_{\mu\nu}=\delta_{\mu\nu}\dot{\Rb}_I$, $\mu \in I$. Force from the band-energy part, the first line in \equ{eq:force}, for atom $I$ can be expressed as
\begin{equation}
\Fb_I = -\textrm{Tr}_I (\rho \dH-\rho^e \dS) + \text{c.c.},
\end{equation}
which is, besides compact, useful in implementation. The density-matrix formulation introduced here is particularly useful in electronic structure analysis, discussed in the following section.

\section{Simplistic electronic structure analysis}
One great benefit of tight-binding is the ease in analyzing the electronic structure. In this section we present selected analysis tools, some old and renowned, others casual but intuitive. Other simple tools for chemical analysis of bonding can be found from Ref.~\onlinecite{mayer_book_03}.

\subsection{Partial Mulliken Populations}
The Mulliken population on atom $I$,
\begin{equation}
q_I=\textrm{Tr}_I (\tilde{\rho} S)=\sum_{\mu \in I} \sum_\nu \tilde{\rho}_{\mu\nu}S_{\nu\mu} =\sum_{\mu \in I} (\tilde{\rho}S)_{\mu\mu},
\end{equation}
is easy to partition into smaller pieces. Population of a single orbital $\mu$ is
\begin{equation}
q_{(\mu)}=\sum_\nu \tilde{\rho}_{\mu\nu}S_{\nu\mu}=(\tilde{\rho}S)_{\mu\mu},
\end{equation}
while population on atom $I$ due to eigenstate $\psi_a$ alone is
\begin{equation}
q_{I,a}=\textrm{Tr}_I (\tilde{\rho}^a S)=\sum_{\mu \in I} \sum_\nu \tilde{\rho}_{\mu\nu}^a S_{\nu\mu},
\end{equation}
so that 
\begin{equation}
\sum_I \left (\sum_a f_a q_{I,a} \right ) = \sum_I (q_I) = N_{el}.
\end{equation}
Population on orbitals of atom $I$ with angular momentum $l$ is, similarly, 
\begin{equation}
q_I^l=\sum_{\mu \in I (l_\mu=l)} (\tilde{\rho}S)_{\mu\mu}.
\end{equation}
The partial Mulliken populations introduced above are simple, but enable surprisingly rich analysis of the electronic structure, as demonstrated below.

\subsection{Analysis Beyond Mulliken Charges}
At this point, after discussing Mulliken population analysis, we comment on the role of wave functions in DFTB. Namely, internally DFTB formalism uses atom resolution for any quantity, and the tight-binding spirit means that the matrix elements $H_{\mu\nu}$ and $S_{\mu\nu}$ are just parameters, nothing more. Nonetheless, the elements $H_{\mu\nu}$ and $S_{\mu\nu}$ are obtained from genuine basis orbitals $\varphi_\mu(\rb)$ using well-defined procedure---these basis orbitals remain constantly available for deeper analysis. The wave functions are
\begin{equation}
\psi_a(\rb)=\sum_\mu c^a_\mu \varphi_\mu(\rb)
\end{equation}
and the total electron density is
\begin{equation} 
n(\rb)=\sum_a f_a |\psi(\rb)|^2=\sum_{\mu\nu} \rho_{\mu\nu} \varphi_{\nu}^*(\rb) \varphi_\mu(\rb),
\end{equation}
awaiting for inspection with tools familiar from DFT. One should, however, use the wave functions only for analysis\cite{yoon_CPC_06}. The formalism itself is better off with Mulliken charges. But for visualization and for gaining understanding this is a useful possibility. This distinguishes DFTB from semiempirical methods, which---in principle---do not possess wave functions but only matrix elements (unless made \emph{ad hoc} by hand).

\subsection{Densities of States}
Mulliken populations provide intuitive tools to inspect electronic structure. Let us first break down the energy spectrum into various components. The complete energy spectrum is given by the density of states (DOS),
\begin{equation} 
\textrm{DOS}(\varepsilon)=\sum_a \delta^\sigma(\varepsilon-\varepsilon_a),
\end{equation}
where $\delta^\sigma(\varepsilon)$ can be either the peaked Dirac delta-function, or some function---such as a Gaussian or a Lorentzian---with broadening parameter $\sigma$. DOS carrying spatial information is the local density of states,
\begin{equation} 
\mathrm{LDOS}(\varepsilon,\rb)=\sum_a \delta^\sigma (\varepsilon-\varepsilon_a)|\psi_a(\rb)|^2,
\end{equation}
with integration over $\int \der^3r$ yielding $\textrm{DOS}(\varepsilon)$. Sometimes
\begin{equation} 
\mathrm{LDOS}(\rb)=\sum_a f'_a |\psi_a(\rb)|^2,
\end{equation}
where $f'_a$ are weights chosen to select states with given energies, as in scanning tunneling microscopy simulations\cite{yin_PRL_09}. Mulliken charges, pertinent to DFTB, yield LDOS with atom resolution,
\begin{equation} 
\mathrm{LDOS}(\varepsilon,I)=\sum_a \delta^\sigma(\varepsilon-\varepsilon_a) q_{I,a},
\end{equation} 
which can be used to project density for group of atoms $\mathcal{R}$ as
\begin{equation}
\mathrm{LDOS}_\mathcal{R}(\varepsilon)=\sum_{I \in \mathcal{R}} \mathrm{LDOS}(\varepsilon,I).
\end{equation}
For instance, if systems consists of surface and adsorbed molecule, we can plot LDOS$_\mathrm{mol}(\varepsilon)$ and LDOS$_\mathrm{surf}(\varepsilon)$ to see how states are distributed; naturally LDOS$_\mathrm{mol}(\varepsilon)+\mathrm{LDOS}_\mathrm{surf}(\varepsilon)=\textrm{DOS}(\varepsilon)$.

Similar recipes apply for projected density of states, where DOS is broken into angular momentum components,
\begin{equation} 
\textrm{PDOS}(\varepsilon,l)=\sum_a \delta^\sigma(\varepsilon-\varepsilon_a) \sum_I q_{I,a}^l,
\end{equation}
such that, again $\sum_l \textrm{PDOS}(\varepsilon,l)=\textrm{DOS}(\varepsilon)$.


\subsection{Mayer Bond-Order}
Bond strengths between atoms are invaluable chemical information. Bond order is a dimensionless number attached to the bond between two atoms, counting the differences of electron pairs on bonding and antibonding orbitals; ideally it is one for single, two for double, and three for triple bonds. In principle, any bond strength measure is equally arbitrary; in practice, some measures are better than others. A measure suitable for many purposes in DFTB is Mayer bond-order\cite{mayer_book_03}, defined for bond $IJ$ as
\begin{equation} 
M_{IJ}= \sum_{\mu \in I, \nu \in J} (\tilde{\rho} S)_{\mu \nu} (\tilde{\rho} S)_{\nu \mu}.
\label{eq:mayer}
\end{equation}
The off-diagonal elements of $\tilde{\rho} S$ can be understood as Mulliken overlap populations, counting the number of electrons in the overlap region---the bonding region. It is straightforward, if necessary, to partition \equ{eq:mayer} into pieces, for inspecting angular momenta or eigenstate contributions in bonding. Look at Refs.~\onlinecite{bridgeman_JCS_01} and \onlinecite{mayer_book_03} for further details, and Table~\ref{tab:energetics} for examples of usage.

\renewcommand{\arraystretch}{0.625}
\begin{table}[tb]
\caption{Simplistic electronic structure and bonding analysis for selected systems: $C_2H_2$ (triple CC bond), $C_2H_4$ (double CC bond), $C_2H_6$ (single CC bond), benzene, and graphene ($\Gamma$-point calculation with $64$ atoms). We list most energy and bonding measures introduced in the main text (energies in eV).}
\begin{tabular}{l r r r r r}
property & $C_2H_2$ & $C_2H_4$ & $C_2H_6$ & benzene & graphene \\
\hline

$q_H$         & 0.85 & 0.94 & 0.96 & 0.95 & \rule{0pt}{3.0ex} \\
$A_H$         & 1.23 & 0.45 & 0.26 & 0.32 &  \\
$AB_H$        & -2.75 & -3.27 & -3.34 & -3.39 &  \\
$E_{\text{prom},H}$  & 0.97 & 0.41 & 0.25 & 0.30 &  \\
$q_C$         & 4.15 & 4.13 & 4.12 & 4.05 & 4.00 \\
$A_C$         & 5.86 & 5.86 & 5.78 & 6.48 & 6.94 \\
$AB_C$        & -9.16 & -9.74 & -10.45 & -9.74 & -9.78 \\
$E_{\text{prom},C}$  & 5.62 & 5.69 & 5.64 & 6.46 & 6.94 \\
$B_{CH}$      & -8.14 & -7.90 & -7.84 & -7.93 &  \\
$B_{CC}$      & -22.01 & -15.82 & -9.39 & -13.01 & -12.16 \\
$M_{CH}$ & 0.96 & 0.95 & 0.97 & 0.96 &  \\
$M_{CC}$ & 2.96 & 2.02 & 1.01 & 1.42 & 1.25 \\[1ex]
\hline 
\end{tabular}
\label{tab:energetics}
\end{table}

\subsection{Covalent Bond Energy}
Another useful bonding measure is the covalent bond energy, which is not just a dimensionless number but measures bonding directly using energy\cite{bornsen_JPCM_99}.

Let us start by discussing promotion energy. When free atoms coalesce to form molecules, higher energy orbitals get occupied---electrons get \emph{promoted} to higher orbitals. This leads to the natural definition
\begin{equation} 
E_\text{prom}=\sum_\mu (q_{(\mu)}-q_{(\mu)}^{\text{free atom}}) H_{\mu\mu}^0.
\end{equation}
Promotion energy is the price atoms have to pay to prepare themselves for bonding. Noble gas atoms, for instance, cannot bind to other atoms, because the promotion energy is too high due to the large energy gap; any noble gas atom could in principle promote electrons to closest $s$-state, but the gain from bonding compared to the cost in promotion is too small.

\begin{figure}
\begin{center}
\includegraphics[width=8cm]{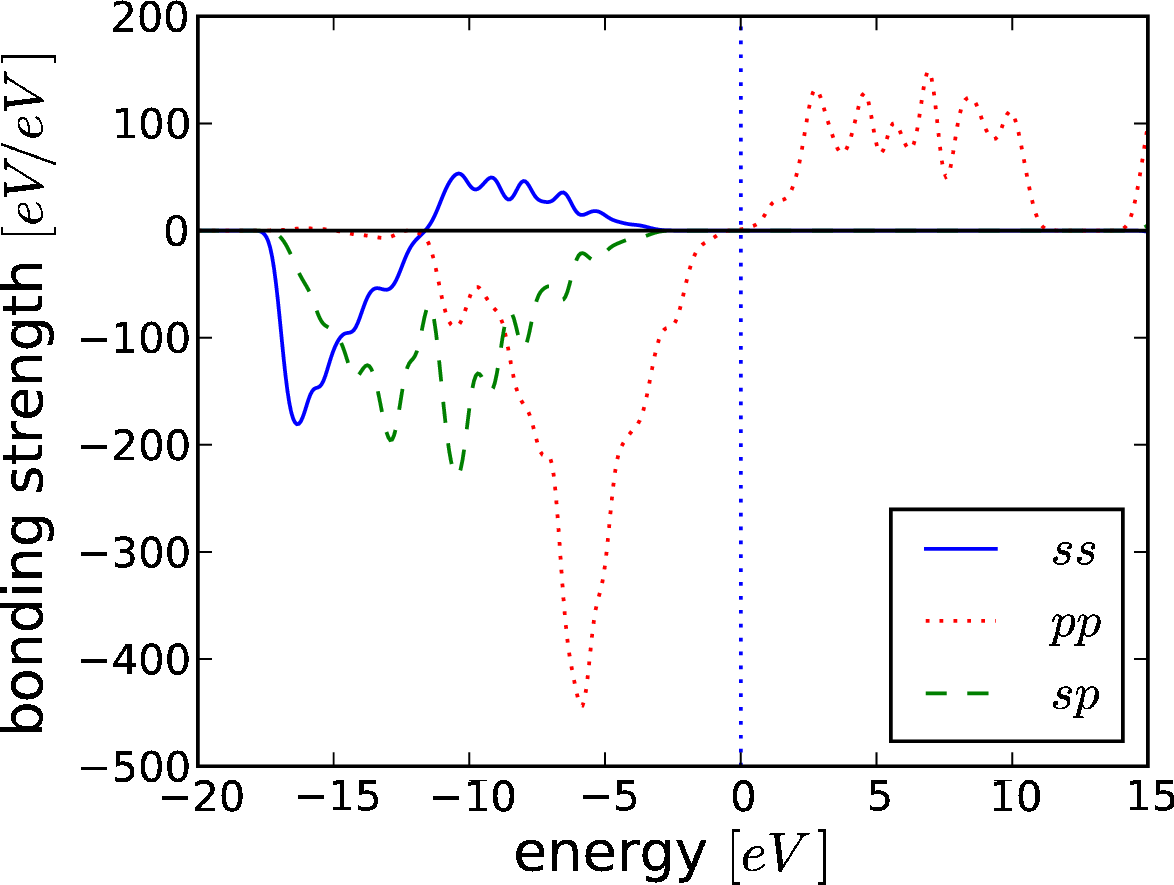}
\end{center}
\caption{(color online) Covalent bonding energy contributions in graphene ($\Gamma$-point calculation with $64$ atoms in unit cell). Both bonding and antibonding $ss$ bonds are occupied, but $sp$ and $pp$ have only bonding contributions. At Fermi-level (zero-energy) the $pp$-bonding (the $\pi$-cloud above and below graphene) transforms into antibonding ($\pi^*$)---hence any addition or removal of electrons weakens the bonds.}

\label{fig:E_cov}
\end{figure}

Covalent bond energy, on the other hand, is the energy reduction from bonding. We define covalent bond energy as\cite{bornsen_JPCM_99}
\begin{equation}
E_\text{cov}=(E_{BS} - E_\text{free atoms}) - E_\text{prom}.
\end{equation}
This definition can be understood as follows. The term $(E_{BS} - E_\text{free atoms})$ is the total gain in band-structure energy as atoms coalesce; but atoms themselves have to pay $E_\text{prom}$, an on-site price that does not enhance binding itself. Subtraction gives the gain the system gets in bond energies as it binds together. More explicitly,
\begin{align}
E_\text{cov}&=E_{BS} - \sum_\mu q_{(\mu)}H_{\mu\mu}^0\\
&=\sum_{\mu\nu} \rho_{\mu\nu}(H^0_ {\mu\nu}-\bar{\varepsilon}_{\mu\nu}S_{\mu\nu}),
\end{align}
where
\begin{equation}
\bar{\varepsilon}_{\mu\nu}=\frac{1}{2}(H_{\mu\mu}^0+H_{\nu\nu}^0).
\end{equation}
This can be resolved with respect to orbital pairs and energy as
\begin{equation}
E_{\text{cov},\mu\nu}(\varepsilon)=\sum_a \delta^\sigma(\varepsilon-\varepsilon_a) \rho^a_{\mu\nu} (H_{\nu\mu}^0-\bar{\varepsilon}_{\nu\mu}S_{\nu\mu}).
\end{equation}
$E_{\text{cov},\mu,\nu}$ can be viewed as the bond strength between orbitals $\mu$ and $\nu$---with strength directly measured in energy; negative energy means bonding and positive energy antibonding contributions. Sum over atom orbitals yields bond strength information for atom pairs
\begin{equation}
E_{\text{cov},IJ}(\varepsilon)=\sum_{\mu \in I} \sum_{\nu \in J} (E_{\text{cov},\mu\nu}(\varepsilon) + \text{c.c.}),
\end{equation}
and sum over angular momentum pairs
\begin{equation}
E_{\text{cov}}^{l_a l_b}(\varepsilon)= \sum_{\mu}^{l_\mu=l_a} \sum_{\nu}^{l_\nu = l_b} (E_{\text{cov},\mu\nu}(\varepsilon) + [\text{c.c.}])
\end{equation}
gives bonding between states with angular momenta $l_a$ and $l_b$ (no c.c. for $l_a=l_b$). For illustration, we plot covalent bonding contributions in graphene in Fig.~\ref{fig:E_cov}.

\subsection{Absolute Atom and Bond Energies}
While Mayer bond-order and $E_\text{cov}$ are general tools for any tight-binding flavor, neither of them take electrostatics or repulsions between atoms into account. Hence, to conclude this section, we introduce a new analysis tool that takes also these contributions into account.

The DFTB energy with subtracted free-atom energies,
\begin{align}
E'=&E-E_\text{free atoms}\\
=&\textrm{Tr} (\rho H^0) + \frac{1}{2}\sum_{IJ}\gamma_{IJ}\Delta q_I \Delta q_J \\
&+ \sum_{I<J} V_{\text{rep}}^{IJ} - \sum_\mu q_{(\mu)}^\textrm{free atom} H_{\mu\mu},
\nonumber
\end{align}
can be rearranged as
\begin{equation} 
E' = \sum_I A_I + \sum_{I<J} B_{IJ},
\label{eq:bind_energy}
\end{equation}
where
\begin{align}
A_I&=\frac{1}{2}\gamma_{II} \Delta q_I^2 + \sum_{\mu \in I} (q_{(\mu)}-q_{(\mu)}^\textrm{free atom})H_{\mu\mu}^0\\
&=\frac{1}{2}\gamma_{II} \Delta q_I^2 + E_{\textrm{prom},I}
\end{align}
and
\begin{align}
\begin{split}
B_{IJ} =& V_{\textrm{rep}}^{IJ} + \gamma_{IJ}\Delta q_I \Delta q_J \\
&+ \sum_{\mu \in I, \nu \in J} \rho_{\nu\mu}\left( H_{\mu\nu}^0-S_{\mu\nu}\bar{\varepsilon}_{\mu\nu} \right) + \text{c.c.}
\end{split}
\end{align}
We call $A_I$ the \emph{absolute atom energy} of atom $I$, and $B_{IJ}$ the \emph{absolute bond energy} of bond $IJ$. Measuring atom energies with $A_I$ and bond energies with $B_{IJ}$ explicitly takes into account all energetics. \equ{eq:bind_energy} can be further simplified into
\begin{equation}
E'=\sum_I \left (A_I+\frac{1}{2}\sum_{J\neq I} B_{IJ} \right ) = \sum_I AB_I,
\end{equation}
where $AB_I$ measures how much atom $I$ contributes to total binding energy---in electrostatic, repulsive, promotive, and bonding sense. For homonuclear crystals the binding energy per atom is directly $AB_I$, and for heteronuclear systems the binding energy per atom (the negative of cohesion energy) is averaged $AB_I$; positive $AB_I$ means atom $I$ would rather be a free atom, even though for charged systems the interpretation of these numbers is more complicated. Visualizing $A_I$, $B_{IJ}$, and $AB_I$ gives a thorough and intuitive measure of energetics; see Table~\ref{tab:energetics} for illustrative examples. Note that $A_I$, $B_{IJ}$, and $AB_I$ come naturally from the exact DFTB energy expression---they are not arbitrary definitions.

\section{Conclusions}
Here ends our journey with DFTB for now. The road up to this point may have been long, but the contents have made it worth the effort: we have given a detailed description of a method to do realistic electronic structure calculations. Especially the transparent chemistry and ease of analysis makes DFTB an appealing method to support DFT simulations. With these features tight-binding methods will certainly remain an invaluable supporting method for years to come.

\section*{Acknowledgements}
One of us (PK) is greatly indebted for Michael Moseler, for introducing molecular simulations in general, and DFTB in particular. Academy of Finland is acknowledged for funding though projects 121701 and 118054. Matti Manninen, Hannu H\"akkinen, and Lars Pastewka are greatly acknowledged for commenting and proof-reading the manu\-script.

\appendix

\section{Calculating the DFT Pseudo-Atom}
\label{app:pseudo_atom}

The pseudo-atom, and also the free atom without the confinement, is calculated with LDA-DFT\cite{perdew_PRB_92}. More recent $xc$-functionals could be used, but they do not improve DFTB parametrizations, whereas LDA provides a fixed level of theory to build foundation. However, better DFT functionals can---and should be---used in the repulsive potential fitting; see Section~\ref{sec:fitting}.

Elements with small atomic numbers are calculated using non-relativistic radial Schr\"odinger equation. But for some elements, such as gold, chemistry is greatly modified by relativistic effects, which have to be included in the atom calculation.

In the four-component Dirac equation with central potential good quantum numbers are energy, total angular momentum $j$, its $z$-component $j_z$, and $-\kappa$ which is the eigenvalue of the operator
\begin{equation}
K=\left ( \begin{array}{cc}
           {\bf \Sigma}\cdot \Lb+1 & 0 \\
0 & -{\bf \Sigma}\cdot \Lb -1\\
          \end{array}
\right ),
\end{equation}
where $\Lb$ is the orbital angular momentum operator and the components of $4\times 4$ relativistic spin-matrix ${\bf \Sigma}$ are 
\begin{equation}
\Sigma_k=\left ( \begin{array}{cc}
           \sigma_k& 0 \\
0 & \sigma_k\\
          \end{array}
\right ),
\end{equation}
with $2\times 2$ Pauli spin-matrices $\sigma_k$. Remember that angular momentum $l$ is \emph{not} a good quantum number; the upper and lower two components are separately eigenstates of $\Lb^2$ with different angular momenta, and coupled by spin-orbit interaction. In other words, a given $l$ (that we are interested in) appears in $4$-component spinors states with different $j$. 

The intention is to include relativistic effects from the Dirac equation, but still use the familiar non-relativistic machinery. This can be achieved by ignoring the spin-orbit interaction, decoupling upper and lower components. By considering only the upper components as a non-relativistic limit, $l$ becomes again a good quantum number. The radial equation transforms into the scalar-relativistic equation\cite{koelling_JPC_77,martin_book}
\begin{align}
\nonumber
&\frac{\der^2R_{nl}(r)}{\der r^2} - \left ( \frac{l(l+1)}{r^2} + 2M(r)(V_s(r)-\varepsilon_{nl}) \right ) R_{nl}(r) \\
&-\frac{1}{M(r)} \frac{\der M(r)}{\der r} \left ( \frac{\der R_{nl}(r)}{\der r} + \langle \kappa \rangle \frac{R_{nl}(r)}{r} \right ) = 0.
\end{align}
Here $\alpha=1/137.036$ is the fine structure constant,
\begin{equation}
V_s(r)=-\frac{Z}{r} + V_H(r) + V_{xc}^\text{LDA}(r) + V_\textrm{conf}(r),
\end{equation}
with or without the confinement, and we defined
\begin{equation}
M(r)=1+\frac{\alpha^2}{2}[\varepsilon_{nl}-V_s(r)].
\end{equation}
The reminiscent of the lower two components of Dirac equation is $\langle \kappa \rangle$, which is the quantum number $\kappa$ averaged over states with different $j$, using the degeneracy weights $j(j+1)$; a straightforward calculation gives $\langle \kappa \rangle=-1$. 

Summarizing shortly, for given $l$ the potential in the radial equation is the weighted average of the potentials in full Dirac theory, with ignored spin-orbit interaction. This scalar-relativistic treatment is a standard trick, and is, for instance, used routinely for generating DFT pseudo-potentials\cite{martin_book}.

The pseudo-atom calculation, as described, yields the localized basis orbitals we use to calculate the matrix elements. Our conventions for the real angular part $\tilde{Y}_\mu(\theta,\varphi)$ of the orbitals $\varphi_\mu(\rb)=R_\mu(r)\tilde{Y}_\mu(\theta,\varphi)$ are shown in Table~\ref{tab:spherical_functions}. The sign of $R_\mu(r)$ is chosen, as usual, such that the first antinode is positive.

\renewcommand{\arraystretch}{1.85}

\begin{table*}
\caption{Spherical functions, normalized to unity ($\int \tilde{Y}(\theta,\varphi) \sin \theta \der \theta \der \varphi=1$) and obtained from spherical harmonics as $\tilde{Y} \propto Y_{lm}\pm Y_{lm}^*$. Note that angular momentum $l$ remains a good quantum number for all states, but magnetic quantum number $m$ remains a good quantum number only for $s$-, $p_z$-, and $d_{3z^2-r^2}$-states.} 
\begin{center}
\begin{tabular}{lll}
visualization& symbol & $\tilde{Y}(\theta,\varphi)$\\
\hline
\vphantom{\rule{0pt}{8mm}} 
\multirow{9}{*}{\includegraphics[width=3.6cm]{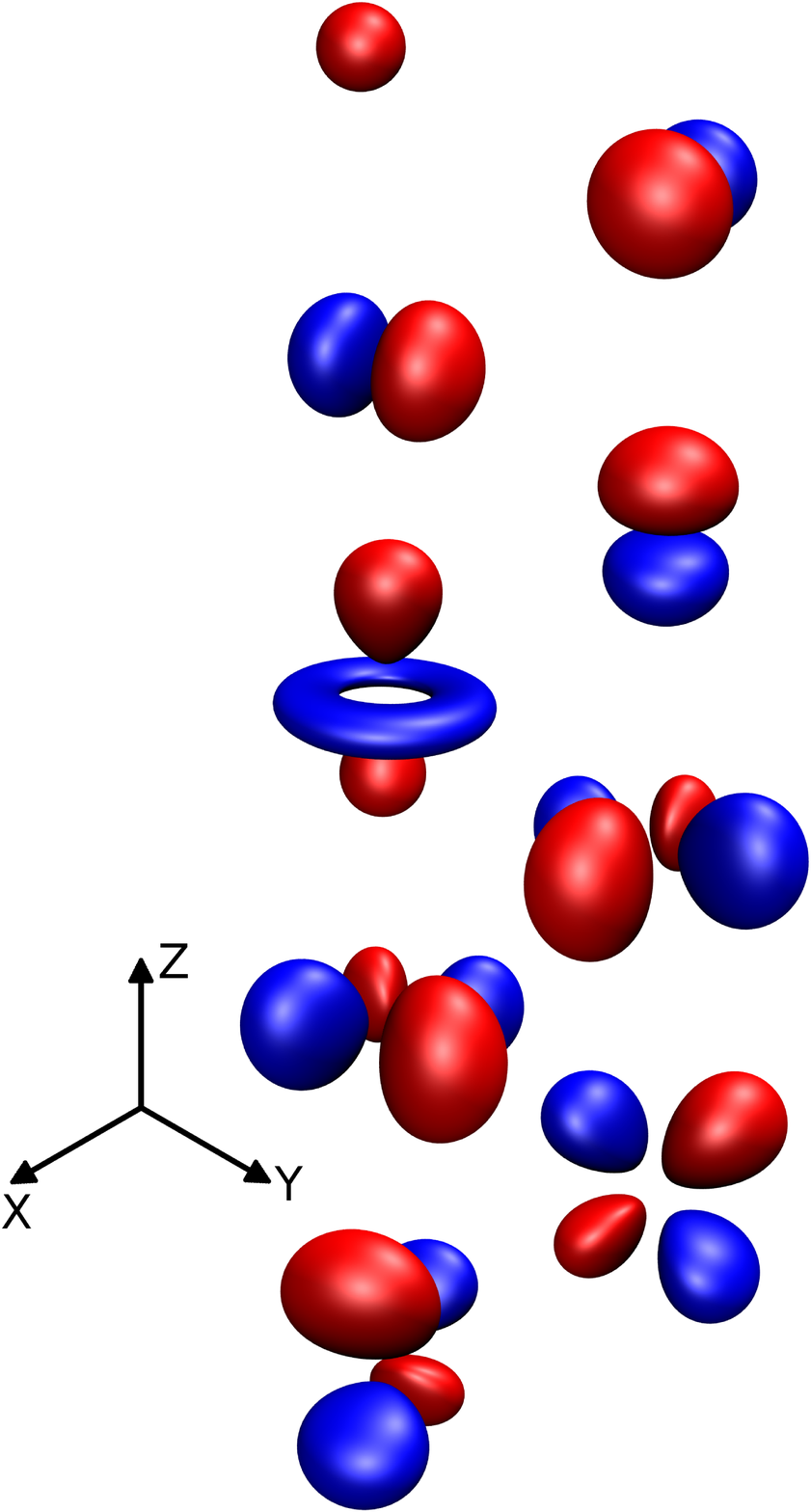}} 
&$s(\theta,\varphi)$             & $\frac{1}{\sqrt{4\pi}}$\\
&$p_x(\theta,\varphi)$           & $\sqrt{\frac{3}{4\pi}} \sin \theta \cos \varphi$\\
&$p_y(\theta,\varphi)$           & $\sqrt{ \frac{3}{4\pi} } \sin \theta \sin \varphi$\\
&$p_z(\theta,\varphi)$           & $\sqrt{ \frac{3}{4\pi} } \cos \theta$\\
&$d_{3z^2-r^2}(\theta,\varphi)$  & $\sqrt{\frac{5}{16\pi} } (3 \cos^2 \theta-1)$\\
&$d_{x^2-y^2}(\theta,\varphi)$   & $\sqrt{ \frac{15}{16\pi} } \sin^2 \theta \cos 2\varphi$\\
&$d_{xy}(\theta,\varphi)$        & $\sqrt{ \frac{15}{16\pi} } \sin^2 \theta \sin 2\varphi$\\
&$d_{yz}(\theta,\varphi)$        & $\sqrt{ \frac{15}{16\pi} } \sin 2\theta \sin \varphi$\\
&$d_{zx}(\theta,\varphi)$        & $\sqrt{ \frac{15}{16\pi} } \sin 2\theta \cos \varphi$\\
\\
\hline
\end{tabular}
\end{center}
\label{tab:spherical_functions}
\end{table*}

\renewcommand{\arraystretch}{1.5}

\section{Slater-Koster transformations}
\label{app:slako}

\begin{figure}[t!]
\begin{center}
\includegraphics[width=6cm]{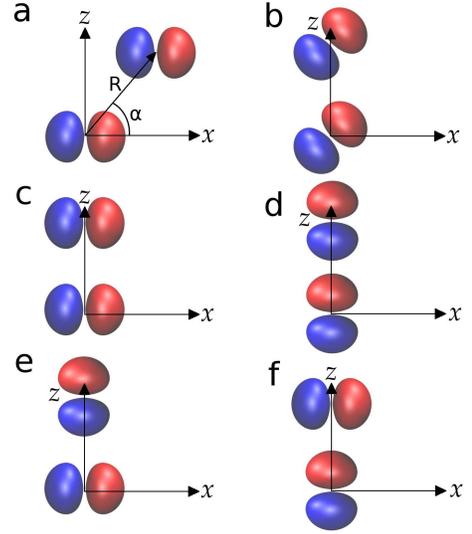}
\end{center}
\caption{(color online) Illustrating overlap integral calculation. (a) Originally one $p_x$ orbital locates at origin, another $p_x$ orbital at $\Rb$. (b) Coordinate system is rotated so that another orbital shifts to $R\hat{z}$; this causes the orbitals in the new coordinate system to become linear combinations of $p_x$ and $p_z$. Hence the overlap can be calculated as the sum of the so-called $S({pp\pi})$-integral in (c), and $S({pp\sigma})$-integral in (d). The integrals in (e) and (f) are zero by symmetry.}
\label{fig:slako_example}
\end{figure}

\renewcommand{\arraystretch}{2.7}
\begin{table*}[tb]
\caption{Calculation of the $10$ Slater-Koster integrals $\tau$. The first orbital (with angular part $\tau_1$) is at origin and the second orbital (with angular part $\tau_2$) at $R\hat{z}$. $\phi_\tau(\theta_1,\theta_2)$ is the function resulting from azimuthal integration of the real spherical harmonics (of Table~\ref{tab:spherical_functions}) $\phi_\tau(\theta_1,\theta_2)=\int_{\varphi=0}^{2\pi} \der \varphi \tilde{Y}_{\tau_1}^*(\theta_1,\varphi)\tilde{Y}_{\tau_2}(\theta_2,\varphi)$, and is used in \equ{eq:overlap_explicit2}. The images in the first row visualize the setup: orbital (\textsf{o}) is centered at origin, and orbital (\textsf{x}) is centered at $R\hat{z}$; shown are wave function isosurfaces where the sign is color-coded, red (light grey) is positive and blue (dark grey) negative.}
\begin{center}
\begin{tabular}{cllll}
& $\tau$ & $\tilde{Y}_{\tau_1}(\theta_1,\varphi)$ & $\tilde{Y}_{\tau_2}(\theta_2,\varphi)$ & $\phi_\tau(\theta_1,\theta_2)$ \\ 
\hline
\vphantom{\rule{0pt}{7mm}} 
\multirow{11}{*}{\includegraphics[width=7.0cm]{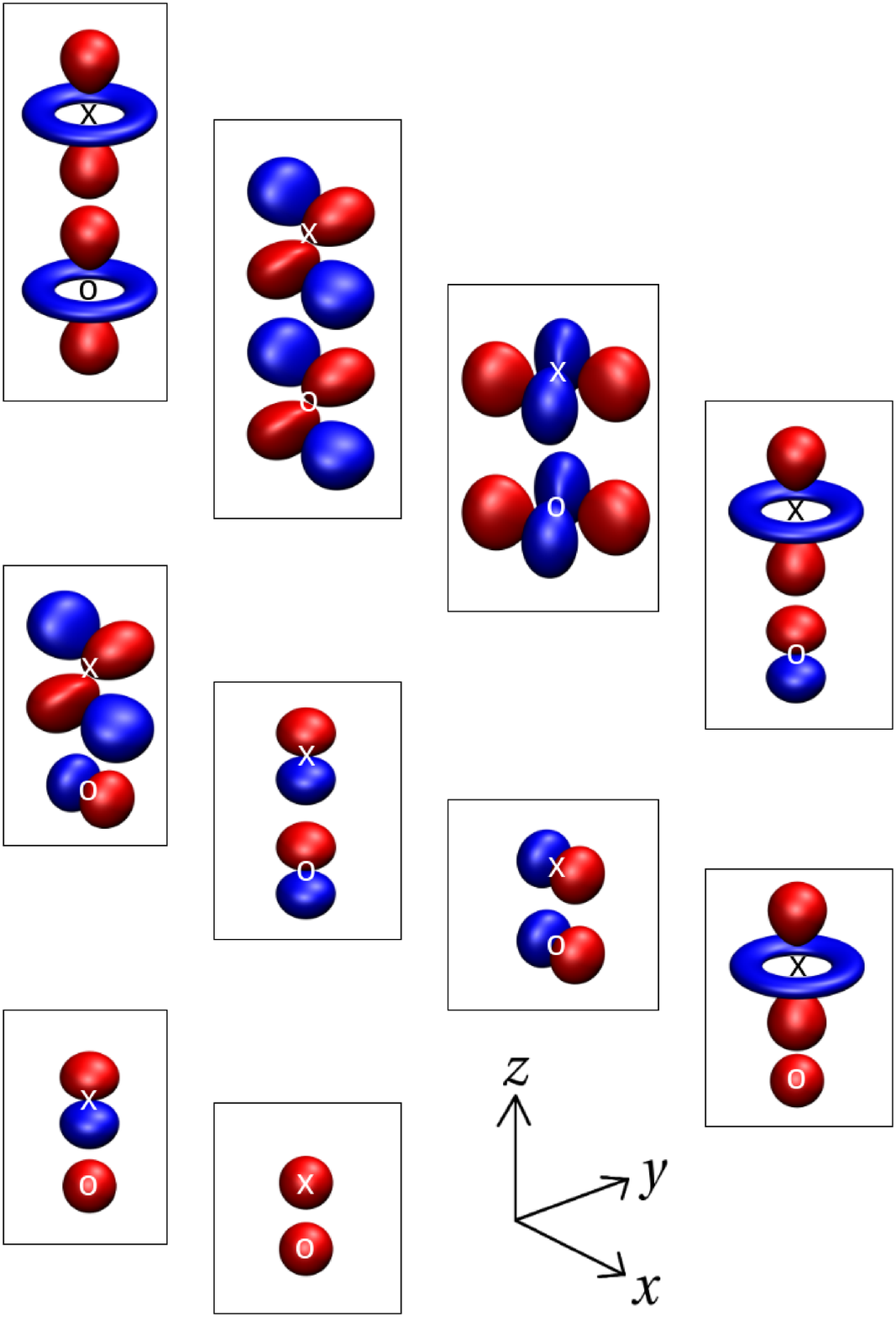}} 
\vphantom{\rule{0pt}{5mm}} \\
&${dd\sigma}$ & $d_{3z^2-r^2}$ & $d_{3z^2-r^2}$    & $\frac{5}{8} (3 \cos^2 \theta_1 -1)(3 \cos^2 \theta_2-1)$  \\
&${dd\pi}   $ & $d_{zx}$       & $d_{zx}$          & $\frac{15}{4} \sin \theta_1 \cos \theta_1 \sin \theta_2 \cos \theta_2$ \\
&${dd\delta}$ & $d_{xy}$       & $d_{xy}$          & $\frac{15}{16} \sin ^2 \theta_1 \sin ^2 \theta_2$ \\
&${pd\sigma}$ & $p_z$          & $d_{3z^2-r^2}$    & $\frac{\sqrt{15}}{4} \cos \theta_1 (3 \cos ^2 \theta_2 -1)$ \\
&${pd\pi}   $ & $p_x$          & $d_{zx}$          & $\frac{\sqrt{45}}{4} \sin \theta_1 \sin \theta_2 \cos \theta_2$ \\
&${pp\sigma}$ & $p_z$          & $p_z$             & $\frac{3}{2} \cos \theta_1 \cos \theta_2$ \\
&${pp\pi}   $ & $p_x$          & $p_x$             & $\frac{3}{4} \sin \theta_1 \sin \theta_2$ \\
&${sd\sigma}$ & $s$            & $d_{3z^3-r^2}$    & $\frac{\sqrt{5}}{4} (3 \cos ^2 \theta_2 -1)$ \\
&${sp\sigma}$ & $s$            & $p_z$             & $\frac{\sqrt{3}}{2} \cos \theta_2$ \\
&${ss\sigma}$ & $s$            & $s$               & $\frac{1}{2}$ \\
\hline
\end{tabular}
\end{center}
\label{tab:integrals}
\end{table*}
\renewcommand{\arraystretch}{1.5}

Consider calculating the overlap integral 
\begin{equation}
S_{xx}=\int p_x(\rb)p_x(\rb-\Rb) \der ^3r 
\end{equation} 
for orbital $p_x$ at origin and another $p_x$ orbital at $\Rb$, as shown in Fig.~\ref{fig:slako_example}a. We rotate the coordinate system passively clockwise, such that the orbital previously at $\Rb$ shifts to $\Rb=R \hat{z}$ in the new coordinate system. Both orbitals become linear combinations of $p_x$ and $p_z$ in the new coordinate system, $p_x \rightarrow p_x \sin \alpha + p_z \cos \alpha$, and the overlap integral becomes a sum of four terms
\begin{align}
\begin{split}
&\int p_x(\rb)p_x(\rb-R\hat{z}) \cdot \sin^2 \alpha  \indent \text{(Fig.~\ref{fig:slako_example}c)}\\
+&\int p_z(\rb)p_z(\rb-R\hat{z}) \cdot \cos^2 \alpha \indent \text{(Fig.~\ref{fig:slako_example}d)}\\
+&\int p_x(\rb)p_z(\rb-R\hat{z}) \cdot \cos \alpha \sin \alpha \indent \text{(Fig.~\ref{fig:slako_example}e)}\\
+&\int p_z(\rb)p_x(\rb-R\hat{z}) \cdot \cos \alpha \sin \alpha \indent \text{(Fig.~\ref{fig:slako_example}f)}.
\end{split}
\end{align}
The last two integrals are zero by symmetry, but the first two terms can be written as 
\begin{equation}
S_{xx} = x^2 S({pp\sigma}) + (1-x^2) S({pp\pi}),
\label{eq:example_sk_rule}
\end{equation}
where $x=\cos \alpha$ is the direction cosine of $\Rb$. For simplicity we assumed $y=0$, but the equation above applies also for $y \neq 0$ (using hindsight we wrote $(1-x^2)$ instead of $z^2$). The integrals
\begin{align}
\begin{split}
S({pp\sigma})&=\int p_z(\rb)p_z(\rb-R\hat{z}) \der^3 r \\
S({pp\pi})&=\int p_x(\rb)p_x(\rb-R\hat{z}) \der^3 r
\end{split}
\end{align}
are called Slater-Koster integrals and \equ{eq:example_sk_rule} is called the Slater-Koster transformation rule for the given orbital pair (orbitals may have different radial parts; the notation $S_{\mu\nu}(pp\sigma)$ stands for radial functions $R_\mu(r)$ and $R_\nu(r)$ in the basis functions $\mu$ and $\nu$). Similar reasoning can be applied for other combinations of $p$-orbitals as well---they all reduce to Slater-Koster transformation rules involving $S({pp\sigma})$ and $S({pp\pi})$ integrals alone. This means that only two integrals with a fixed $R$ is needed for \emph{all} overlaps between any two $p$-orbitals from a given element pair.

Finally, it turns out that $10$ Slater-Koster integrals, labeled $dd\sigma$, $dd\pi$, $dd\delta$, $pd\sigma$, $pd\pi$, $pp\sigma$, $pp\pi$, $sd\sigma$, $sp\sigma$, and  $ss\sigma$, are needed to transform all $s$-, $p$-, and $d$- matrix elements. The last symbol in the notation, $\sigma$, $\pi$, or $\delta$, refers to the angular momentum around the symmetry axis, and is generalized from the atomic notation $s$, $p$, $d$ for $l=0,1,2$.

Table~\ref{tab:integrals} shows how to select the angular parts for calculating these $10$ Slater-Koster integrals. The integrals can be obtained in many ways; here we used our setup with the first orbital at the origin and the second orbital at $R\hat{z}$. This means that we set the direction cosines $z=1$ and $x=y=0$ in Table~\ref{tab:slako_transformations}, and chose orbital pairs accordingly.

Finally, Table~\ref{tab:slako_transformations} shows the rest of the Slater-Koster transformations. The overlap $S_{\mu\nu}$ between orbitals $\varphi_\mu$ at $\Rb_\mu=0$ and $\varphi_\nu$ at $\Rb_\nu=\Rb$ is the sum 
\begin{align}
S_{\mu\nu}&=\sum_\tau c_{\tau} S_{\mu\nu}(\tau), 
\end{align}
for the pertinent Slater-Koster integrals $\tau$ (at most three). The gradients of the matrix elements come directly from the above expression by chain rule: Slater-Koster integrals $S_{\mu\nu}(\tau)$ depend only on $R$ and the coefficients $c_{\tau}$ only on $\hat{R}$.

For $9$ orbitals (one $s$-, three $p$-, and five $d$-orbitals) $81$ transformations are required, whereas only $29$ are in Table~\ref{tab:slako_transformations}. Transformations with an asterisk can be manipulated to yield another $16$ transformations and the remaining ones can be obtained by inversion, which effectively changes the order of the orbitals. This inversion changes the sign of the integral according to orbitals' angular momenta $l_\mu$ and $l_\nu$,
\begin{equation}
S_{\mu\nu}(\tau)=S_{\nu\mu}(\tau)\cdot (-1)^{l_\mu+l_\nu},
\end{equation}
because orbital parity is $(-1)^l$.

The discussion here concentrated only on overlap, but the Slater-Koster transformations apply equally for Hamiltonian matrix elements.

\def\half{ \frac{1}{2} }
\def\vali{ \hspace{0.3cm} }
\def\xpy{\alpha} 
\def\xmy{\beta} 

\renewcommand{\arraystretch}{0.85}
\begin{table*}[tb]
\caption{Slater-Koster transformations for $s$-, $p$-, and $d$-orbitals, as first published in Ref.~\onlinecite{slater_PR_54}. To shorten the notation we used $\xpy=x^2+y^2$ and $\xmy=x^2-y^2$. Here $x$, $y$, and $z$ are the direction cosines of $\hat{\Rb}$, with $x^2+y^2+z^2=1$. 
Missing transformations are obtained by manipulating transformations having an asterisk ($*$) in the third column, or by inversion. Here $m$ is $\varphi_\mu$'s angular momentum and $n$ is $\varphi_\nu$'s angular momentum.}
\begin{center}
\begin{tabular}{l l l l l l}
$\mu$ at 0    & $\nu$ at $\Rb$            &  & $c_{mn\sigma}$ & $c_{mn\pi}$ & $c_{mn\delta}$ \\
\hline
$s$             & $s$               &  & $1$                                       & &\\
$s$             & $p_x$             &* & $x$                                       & &\\
$s$             & $d_{xy}$          &* & $ \sqrt{3} xy$                            & &\\
$s$             & $d_{x^2-y^2}$     &  & $ \half \sqrt{3} \xmy$                    & &\\
$s$             & $d_{3z^2-r^2}$    &  & $ z^2-\half \xpy$                         & &\\
$p_x$           & $p_x$             &* & $x^2$                                     & $1-x^2$\\
$p_x$           & $p_y$             &* & $xy$                                      & $-xy$ &\\
$p_x$           & $p_z$             &* & $xz$                                      & $-xz$ &\\
$p_x$           & $d_{xy}$          &  & $\sqrt{3} x^2 y$                          & $y(1-2x^2) $\\
$p_x$           & $d_{yz}$          &  & $\sqrt{3}xyz$                             & $-2xyz$\\
$p_x$           & $d_{zx}$          &  & $\sqrt{3}x^2z$                            & $z(1-2x^2)$\\
$p_x$           & $d_{x^2-y^2}$     &  & $\half \sqrt{3} x\xmy$                    & $x(1-\xmy)$ &\\
$p_y$           & $d_{x^2-y^2}$     &  & $\half \sqrt{3} y\xmy$                    & $-y(1+\xmy)$ &\\
$p_z$           & $d_{x^2-y^2}$     &  & $\half \sqrt{3} z\xmy$                    & $-z\xmy$                  &\\
$p_x$           & $d_{3z^2-r^2}$    &  & $x(z^2-\half \xpy)$                       & $-\sqrt{3} xz^2$          &\\
$p_y$           & $d_{3z^2-r^2}$    &  & $y(z^2-\half \xpy)$                       & $-\sqrt{3}yz^2$           &\\
$p_z$           & $d_{3z^2-r^2}$    &  & $z(z^2-\half \xpy)$                       & $\sqrt{3} z\xpy$          &\\
$d_{xy}$        & $d_{xy}$          &* & $3 x^2y^2$                                & $\xpy-4 x^2y^2$        & $z^2+x^2y^2$\\
$d_{xy}$        & $d_{yz}$          &* & $3xy^2z$                                  & $xz(1-4y^2)$              & $xz(y^2-1) $\\
$d_{xy}$        & $d_{zx}$          &* & $3x^2yz$                                  & $yz(1-4x^2)$              & $yz(x^2-1) $\\
$d_{xy}$        & $d_{x^2-y^2}$     &  & $\frac{3}{2}xy\xmy$                       & $-2xy\xmy$                & $\half xy\xmy $\\
$d_{yz}$        & $d_{x^2-y^2}$     &  & $\frac{3}{2}yz\xmy$                       & $-yz(1+2\xmy)$            & $yz(1+\half \xmy) $\\
$d_{zx}$        & $d_{x^2-y^2}$     &  & $\frac{3}{2}zx\xmy$                       & $zx(1-2\xmy)$             & $-xz(1-\half \xmy)  $\\
$d_{xy}$        & $d_{3z^2-r^2}$    &  & $\sqrt{3}xy(z^2-\half \xpy)$              & $-2\sqrt{3} xyz^2$        & $\half \sqrt{3} xy (1+z^2)  $\\
$d_{yz}$        & $d_{3z^2-r^2}$    &  & $\sqrt{3}yz(z^2-\half \xpy)$              & $\sqrt{3}yz (\xpy-z^2)$   & $-\half \sqrt{3} yz\xpy $\\
$d_{zx}$        & $d_{3z^2-r^2}$    &  & $\sqrt{3}xz(z^2-\half \xpy)$              & $\sqrt{3} xz(\xpy-z^2)$   & $-\half \sqrt{3} xz\xpy  $\\
$d_{x^2-y^2}$   & $d_{x^2-y^2}$     &  & $\frac{3}{4} \xmy^2$                      & $\xpy-\xmy^2$             & $z^2+\frac{1}{4}\xmy^2 $\\
$d_{x^2-y^2}$   & $d_{3z^2-r^2}$    &  & $\half \sqrt{3} \xmy (z^2-\half \xpy)$    & $-\sqrt{3}z^2\xmy$       & $\frac{1}{4} \sqrt{3} (1+z^2)\xmy$\\
$d_{3z^2-r^2}$   & $d_{3z^2-r^2}$   &  & $(z^2-\half \xpy)^2$                      & $3z^2\xpy$                & $\frac{3}{4}\xpy^2$
\end{tabular}
\end{center}
\label{tab:slako_transformations}
\end{table*}

\section{Calculating the Slater-Koster Integrals}
\label{app:mels}
\subsection{Overlap Integrals}
\label{app:Smels}
We want to calculate the Slater-koster integral 
\begin{equation}
S_{\mu\nu}(\tau)=\langle \varphi_{\mu\tau_1}|\varphi_{\nu\tau_2} \rangle,
\label{eq:simple_S_integral}
\end{equation} 
with 
\begin{equation}
\varphi_{\mu \tau_1}(\rb)=R_\mu(r)\tilde{Y}_{\tau_1}(\theta,\varphi)=R_\mu(r)\Theta_{\tau_1}(\theta) \Phi_{\tau_1}(\varphi)
\label{eq:basisfunc}
\end{equation}
and
\begin{equation}
\varphi_{\nu\tau_2}(\rb)=R_\nu(r)\tilde{Y}_{\tau_2}(\theta,\varphi)=R_\nu(r)\Theta_{\tau_2}(\theta) \Phi_{\tau_2}(\varphi),
\label{eq:basisfunc2}
\end{equation}
where $R(r)$ is the radial function, and the angular parts $\tilde{Y}_{\tau_i}(\theta,\varphi)$ are chosen from Table~\ref{tab:integrals} and depend on the Slater-Koster integral $\tau$ in question. Like in Appendix~\ref{app:slako}, we choose $\varphi_\mu$ to be at the origin, and $\varphi_\nu$ to be at $\Rb=R \hat{z}$.

Explicitly,
\begin{align}
\begin{split}
S_{\mu\nu}(\tau)=\int \der^3 r &[R_\mu(r_1) \Theta_{\tau_1}(\theta_1) \Phi_{\tau_1}(\varphi_1)] \\
\times &[R_\nu(r_2) \Theta_{\tau_2}(\theta_2) \Phi_{\tau_2}(\varphi_2)],
\end{split}
\end{align}
where $\rb_1=\rb$ and $\rb_2=\rb-R \hat{z}$. Switching to cylindrical coordinates we get
\begin{align}
\begin{split}
 S_{\mu\nu}(\tau)=&\int\int \der z\rho \der \rho R_\mu(r_1)R_\nu(r_2)  \\ 
&\times \Theta_{\tau_1}(\theta_1)\Theta_{\tau_2}(\theta_2) \int_{\varphi =0}^{2\pi} \Phi_{\tau_1}(\varphi_1)\Phi_{\tau_2}(\varphi_2) \der \varphi,
\end{split}
\label{eq:overlap_explicit1}
\end{align}
and we see that since $\hat{z}$-axis remains the symmetry axis, the $\varphi$-integration can be done analytically. The second line in \equ{eq:overlap_explicit1} becomes an analytical expression $\phi_\tau(\theta_1,\theta_2)$, and is given in Table~\ref{tab:integrals}. Note here that $r$ is a spherical coordinate, whereas $\rho$ is the distance from a $\hat{z}$-axis in cylindrical coordinates. We are left with
\begin{align}
S_{\mu\nu}(\tau)=\int \int \der z \rho \der \rho R_\mu(r_1)R_\nu(r_2) \phi_\tau(\theta_1,\theta_2),
\label{eq:overlap_explicit2}
\end{align}
a two-dimensional integral to be integrated numerically.

\subsection{Hamiltonian Integrals}
\label{app:Hmels}
The calculation of the Slater-Koster Hamiltonian matrix elements 
\begin{equation}
H_{\mu\nu}(\tau)=\langle \varphi_{\mu\tau_1}|H^0|\varphi_{\nu\tau_2} \rangle
\label{eq:simple_H_integral}
\end{equation} 
is mostly similar to overlap. The potentials $V_{s,I}[n_{0,I}](\rb)$ in the Hamiltonian 
\begin{equation}
H^0=-\frac{1}{2} \nabla^2 + V_{s,I}[n_{0,I}](\rb) + V_{s,J}[n_{0,J}](\rb),
\end{equation}
with $\mu \in I$ and $\nu \in J$, are approximated as
\begin{equation}
V_{s,I}[n_{0,I}](r) \approx V_{s,I}^\text{conf}(r)-V_{\textrm{conf},I}(r),
\end{equation}
where $V_{s,I}^\text{conf}(r)$ is the self-consistent effective potential from the \emph{confined} pseudo-atom, but without the confining potential. The reasoning behind this is that while $V_\textrm{conf}(r)$ yields the pseudo-atom and the pseudo-atomic orbitals, the potential $V_s[n_0](\rb)$ in $H^0$ should be the approximation to the true crystal potential, and should not be augmented by confinements anymore. The Hamiltonian becomes
\begin{align}
\begin{split}
H^0=-\frac{1}{2} \nabla^2 &+ V_{s,I}^\text{conf}(\rb) - V_\text{conf,I}(\rb)\\
&+ V_{s,J}^\text{conf}(\rb) - V_\text{conf,J}(\rb)
\end{split}
\end{align}
and the matrix element
\begin{align}
\label{eq:hamiltonian_two_times1}
H_{\mu\nu}(\tau)=&\varepsilon_\mu^\text{conf} S_{\mu\nu}(\tau)\nonumber\\
&+ \langle \varphi_{\mu\tau_1}|V_{s,J}^\textrm{conf} 
-V_\textrm{conf,I} - V_\textrm{conf,J}|\varphi_{\nu\tau_2} \rangle \\
\label{eq:hamiltonian_two_times2}
=&\varepsilon_\nu^\text{conf} S_{\mu\nu}(\tau) \nonumber\\
&+ \langle \varphi_{\mu\tau_1}|V_{s,I}^\textrm{conf} 
-V_\textrm{conf,I} - V_\textrm{conf,J}|\varphi_{\nu\tau_2} \rangle,
\end{align}
depending whether we operate left with $-\nabla^2/2 + V_{s,I}^\text{conf}$ or right with $-\nabla^2/2 + V_{s,J}^\text{conf}$; $\varphi_\mu$'s are eigenstates of the confined atoms with eigenvalues $\varepsilon_\mu^\text{conf}$ (including the confinement energy contribution which is then subtracted). The form used in numerical integration is
\begin{align} 
\begin{split}
H_{\mu\nu}(\tau)=&\varepsilon_\mu^\text{conf} S_{\mu\nu}(\tau) \\
&+ \int \int \der z \rho \der \rho R_\mu(r_1)R_\nu(r_2) \phi_\tau(\theta_1,\theta_2) \\
&\times \left[V_{s,I}^\textrm{conf}(r_1)-V_\textrm{conf,I}(r_1) - V_\textrm{conf,J}(r_2) \right].
\end{split}
\label{eq:hamiltonian_integration}
\end{align}
As an internal consistency check, we can operate to $\varphi_\nu$ also directly with $\nabla^2$, which in the end requires just $\der^2 R_\nu(r)/\der r^2$, but gives otherwise similar integration. Comparing the numerical results from this and the two versions of Eqs.~(\ref{eq:hamiltonian_two_times1}) and (\ref{eq:hamiltonian_two_times2}) give way to estimate the accuracy of the numerical integration.

Note that the potential in \equ{eq:hamiltonian_two_times1} diverges as $\rb \rightarrow R\hat{z}$, and the potential in \equ{eq:hamiltonian_two_times1} diverges as $\rb \rightarrow 0$. For this reason we use two-center polar grid, centered at the origin and at $R\hat{z}$, where the two grids are divided by a plane parallel to $xy$-plane, and intersecting with the $\hat{z}$-axis at $\frac{1}{2}R\cdot \hat{z}$.



\end{document}